\documentclass[useAMS,usenatbib]{mnras}

\usepackage{graphicx}
\usepackage{scalefnt}
\usepackage{amssymb}
\usepackage{amsmath}
\usepackage[german]{babel}
\usepackage{hyperref}
\usepackage{balance}
\usepackage{xspace}
\usepackage{chngcntr}

\newcommand{\yr}	    {\ifmmode \mathrm{yr} \else yr\fi\xspace}
\newcommand{\mpc}	{\ifmmode \,\mathrm{Mpc}^{-3} \else \,Mpc$^{-3}$\fi\xspace}
\newcommand{\Msun}	{\ifmmode \,\mathrm M_{\odot} \else $\,\mathrm M_{\odot}$\fi\xspace}
\newcommand{\Zsun}	{\ifmmode \,\mathrm Z_{\odot} \else $\,\mathrm Z_{\odot}$\fi\xspace}
\newcommand{\erg}	{\ifmmode \,\mathrm erg \else erg\fi\xspace}
\newcommand{\Mhalo}	{\ifmmode \,\mathrm M_{\mathrm{halo}} \else
  $\,\mathrm M_{\mathrm{halo}}$\fi\xspace}
\newcommand{\Rvir}	{\ifmmode R_{\mathrm{crit,200}} \else $R_{\rm
    crit,200}$\fi\xspace}
\newcommand{\Mstar}	{\ifmmode {M}_{\star} \else ${M}_{\star}$\fi\xspace}
\newcommand{\Mvir}	{\ifmmode M_{\mathrm{crit,200}} \else $M_{\rm
    crit,200}$\fi\xspace}
\newcommand{\Htwo}	{\ifmmode {\rm H}_{2} \else ${\rm H}_{2}$\fi\xspace}
\newcommand{\nH}	{\ifmmode {n}_{\rm H} \else ${n}_{\rm H}$\fi\xspace}
\newcommand{\cc}	{\ifmmode {\rm cm}^{-3} \else ${\rm cm}^{-3}$\fi\xspace}
\newcommand{\arepo}	{\textsc{arepo}\xspace}

\newcommand{\varSN}	{\texttt{Variable\_SN\_energy}\xspace}
\newcommand{\fixSN}	{\texttt{Fixed\_SN\_energy}\xspace}

\newcommand{\nofb}	{\texttt{No\_FB}\xspace}

\defcitealias{Hu2017}{H17}
\defcitealias{Emerick2019}{E19}

\title[LYRA project I]{LYRA I: Simulating the multi-phase ISM of a dwarf galaxy with variable energy supernovae from individual stars}
\author[T. A. Gutcke et al.]{Thales
  A. Gutcke$^1$\thanks{thales@mpa-garching.mpg.de},
  R\"{u}diger Pakmor$^1$, Thorsten Naab$^1$ and
  Volker Springel$^1$\\
$^{1}$Max-Planck-Institut f\"ur Astrophysik, Karl-Schwarzschild-Str. 1, D-85748, Garching, Germany}

\begin{document}
\pagerange{\pageref{firstpage}--\pageref{lastpage}} \pubyear{---}
\maketitle
\label{firstpage}
\begin{abstract}
We introduce the LYRA project, a new high resolution galaxy formation model built within the framework of the cosmological hydro-dynamical moving mesh code \arepo. The model resolves the multi-phase interstellar medium down to 10\,K. It forms individual stars sampled from the initial mass function (IMF), and tracks their lifetimes and death pathways individually. Single supernova (SN) blast waves with variable energy are followed within the hydrodynamic calculation to interact with the surrounding interstellar medium (ISM). In this paper, we present the methods and apply the model to a $10^{10}\Msun$ isolated halo. We demonstrate that the majority of supernovae are Sedov-resolved at our fiducial gas mass resolution of $4\,\Msun$. We show that our SN feedback prescription self-consistently produces a hot phase within the ISM that drives significant outflows, reduces the gas density and suppresses star formation. Clustered SN play a major role in enhancing the effectiveness of feedback, because the majority of explosions occur in low density material. {Accounting for variable SN energy allows the feedback to respond directly to stellar evolution. We show that the ISM is sensitive to the spatially distributed energy deposition. It strongly affects the outflow behaviour, reducing the mass loading by a factor of $2-3$, thus allowing the galaxy to retain a higher fraction of mass and metals.} LYRA makes it possible to use a comprehensive multi-physics ISM model directly in cosmological (zoom) simulations of dwarf and higher mass galaxies.
\end{abstract}
\section{Introduction}
%
%
%
In the current paradigm of galaxy formation in a $\Lambda$CDM universe, the non-linear baryonic physics involved in structure formation and galaxy build-up is an active area of research. Not only may baryons play an important role in shaping dark matter halo profiles \citep[e.g.][]{El-Zant2001, Weinberg2002, Read2005, Pontzen2012}, but the details of the star formation process and the ensuing injection of energetic feedback from SNe are decisive to the formation and evolution of luminous galaxies \citep{Larson1974, Saito1979, White1978, Dekel1986, MacLow1999}. In the centers of dark matter halos we expect gas to collapse into molecular clouds, cool radiatively and fragment to form stars. Stars are expected to emit radiation and in some cases die as supernovae (SNe), ejecting energy and enriching the interstellar medium (ISM) with metals. The metals, in turn, enhance the cooling process and the cycle repeats. 

Cosmological, hydrodynamic simulations of galaxy formation have been instrumental in developing our present understanding of the baryon cycle in galaxies. However, modelling the relevant processes has proven to be non-trivial, leaving many uncertainties unresolved. Dwarf galaxies are particularly interesting laboratories to study galaxy formation. Interest in these systems truly began to surge after the Sloan Digital Sky Survey discovered a large population of ultra-faint and classical dwarfs in and around the Milky Way \citep[e.g.][]{Willman2005, Belokurov2006}, see \citet{Simon2019} for a review. Because they are simultaneously low mass and dark matter-dominated, dwarf galaxies are a critical testing ground for dark matter models. Additionally, they are extreme environments in terms of mass, metallicity and size. Their shallow potential wells make them highly sensitive to reionization and feedback, while making them also easier to model computationally than larger systems. Thus, the past decade has given rise to many published studies achieving substantial progress in our theoretical understanding of dwarf galaxies and their place in galaxy evolution \citep[e.g.][]{Governato2010, Governato2015, Sawala2011, Munshi2013, Simpson2013, Trujillo-Gomez2015, Onorbe2015, Maccio2017, Frings2017}. 

The ISM in such cosmological simulations has often been modelled with  effective (sub-grid) models, for example following \citet{Springel2003}. This prevents the self-consistent formation of cold ($T<10^4$\,K) and dense ($\nH>1$\,\cc) gas, primarily for computational reasons. An ISM modelled in such a way does not form giant molecular clouds (GMCs), which are known to be the sites of star formation. Without these structures, the local conditions into which stars form cannot be resolved. Moreover, this low temperature ISM may be critical for resolving the effect of feedback especially in low-mass systems that are very sensitive to individual feedback events.

Recently, it has become computationally feasible to run high resolution cosmological zoom-in simulations of field dwarfs with a resolved ISM that includes cooling below $10^4\,$K. For example, a simulation project that has made significant progress simulating cosmological dwarf galaxies with a multi-phase ISM is \citet{Smith2019} using the model developed in \citet{Smith2018}. The authors present five dwarfs in the
halo mass range $2.5-6\times10^9\Msun$ that they run to $z=4$ with a baryonic mass resolution of $\sim300\,\Msun$.
Using single stellar population star particles and a SN feedback prescription that injects momentum instead of thermal energy when the SN is estimated to be under-resolved, they conclude that other forms of stellar feedback beyond SNe alone are
necessary to explain the low SF efficiencies observed in dwarf
galaxies.

The EDGE project \citep{Agertz2020} presented a $10^9$\,\Mhalo dwarf galaxy resolving down to $\sim300\,M_{\odot}$ single stellar populations for which they resolve the multi-phase ISM, individual SN and radiation via radiation transfer \citep[with RAMSES-RT,][]{Rosdahl2013} from young stars in a cosmological context. They conclude that the galaxy-averaged stellar metallicity is a sensitive indicator able to discriminate between feedback models.
They also show that the radiation significantly
affects the growth of the galaxy. Runs with radiation have lower
stellar masses by about an order of magnitude. They attribute this to
the radiative heating of much of the gas to $\sim10^4$\,K, preventing it
from cooling and condensing to densities above their SF threshold.

The FIRE-2 galaxy formation model
presented in \citet{Hopkins2018} introduced a method to help resolve SN blastwaves by adding the unresolved momentum explicitly, while also accounting for low temperature cooling. Using this model, \citet{Wheeler2019} simulate three high resolution cosmological dwarf galaxies at a mass resolution of $\sim30\,\Msun$ with
$z=0$ virial masses of $\sim2-10\times10^9\Msun$, testing their survival throughout reionization. In most cases SF is quenched before $z=10$ due to reionization. The more massive systems have residual SF until $z\sim2-4$ that is kept going due to gas being self-shielded. 

Concurrently, non-cosmological simulations of individual clouds and stratified disks have made great strides to develop more accurate models of the ISM and capture baryonic feedback effects on a much smaller scale \citep[e.g.][]{Glover2007, Walch2015a, Walch2015b, Kim2015, Girichidis2016, Simpson2016, Kim2017, SmithR2020, Hirai2020}. A critical insight into the process of SN feedback that has emerged from these efforts is that the density into which a SN shock expands is crucial to its effect on the ISM and, thus, the galaxy as a whole \citep[see][for a discussion]{Naab2017}. SNe in dense environments have a negligible effect both dynamically and thermally \citep[e.g. see][]{Gatto2015}. The largest feedback effect (preventing star formation) is caused by SNe expanding into lower density regions and heating a significant fraction (depending on the simulation, 50-90 \% of the volume of the ISM) to $\sim10^6$\,K. According to \citet{Gatto2015}, this is only achieved numerically if the SN is injected thermally. Momentum injection accounts for the unresolved dynamical contribution of the SNe but does not produce a hot phase that is essential for driving outflows and preventing star formation \citep{Hu2019,Steinwandel2020}.
Additionally, SNe in dense gas are more difficult to resolve numerically, since their Sedov radii are small ($\sim 1$\,pc) and the onset of the momentum conserving phase occurs much more quickly than for SNe expanding into more diffuse gas.
Fortunately for simulations, a large fraction of stars are born in star clusters \citep{Lada2003} which in turn leads to the SNe being clustered as well. Thus, the majority of SN explode into the evacuated regions of previous SNe, making them easier to resolve numerically while at the same time enhancing their feedback impact.

Recently, these improved ISM and feedback models are being put to the test on isolated dwarf galaxy simulations using particle-based \citep{Hu2016, Hu2017, Lahen2019, Lahen2020} and Eulerian grid codes \citep{Emerick2019, Emerick2020}. These models are able to begin disentangling the relative contribution of different feedback processes such as SNe, stellar winds and radiation for young stars while comparing the results to integrated galaxy properties. \citet{Hu2017} considered feedback from individual massive stars and a chemical network allowing for the formation of a multi-phase ISM with a resolved cold, warm, and -- of particular importance -- hot phase \citep[see e.g.][]{Steinwandel2020}. This is also the basic model for the GRIFFIN project \citep{Lahen2020} which studies a merger driven dwarf galaxy starburst with individual massive stars at sub-parsec resolution, and investigates the self-consistent formation of star clusters.
{\citet{Smith2020} also use an isolated dwarf galaxy to test the effects of a sampled IMF relative to using an integrated IMF prescription. Thus, there is a growing body of work showing that the details of the SF and feedback processes have a significant impact on the resulting galaxy properties.}

In this work we introduce LYRA, a novel galaxy formation physics model for the moving mesh code \arepo with low temperature cooling and resolved feedback from individual stars. The aim of this new model is to attain sufficiently high mass resolution that individual blastwaves are resolved, thereby eliminating many of the uncertainties associated with the sub-grid modelling of SN energy injection. This, in turn, enables the model to capture the interaction of the SN blastwaves with the ISM and the subsequent formation of outflows. 
Simultaneously, we make modelling choices that allow the use of this model in cosmological zoom-in initial conditions of dwarf galaxies. This paper details the model and presents the first tests using an isolated { low mass} dwarf galaxy. In Paper II (Gutcke et al. in prep) we will analyze cosmological zoom-in simulations using the LYRA model.

The paper is organized as follows: Section~\ref{sec:model} describes the implementation details of the LYRA model. Section~\ref{sec:SNboxes} examines how well individual SN are resolved, while Section~\ref{sec:isolated} applies the LYRA model to an isolated dwarf galaxy and details its feedback properties. In Section~\ref{sec:discussion} we compare our results to previous work and discuss limitations due to missing physics, while Sections~\ref{sec:conclusion} summarizes our conclusions. Appendix~\ref{sec:appendix} presents convergence and parameter tests.
%
\section{The LYRA model}
\label{sec:model}
\subsection{Framework}
Our model is built into the moving-mesh magnetohydrodynamics (MHD) code \arepo \citep{Springel2010,Pakmor2016,Weinberger2020}. It solves the magneto-hydrodynamic (MHD) equations on an unstructured Voronoi mesh that adapts itself to the gas flow using a second order finite volume scheme. In this way \arepo combines the accuracy of Eulerian finite volume schemes, in particular their superb shock capturing abilities, with  advantages offered by Lagrangian codes, such as their excellent spatial adaptivity and accuracy for high bulk flow velocities.

In the quasi-Lagrangian mode of \arepo, the mesh-generating points that are used to construct the Voronoi tesselation follow the gas motions, with a small correction to keep the mesh regular \citep{Vogelsberger2013}. Owing to the non-linear nature of the MHD equations, cells can still accumulate or lose mass over time. Therefore, to keep the mass of the cells within a factor of two of our target mass resolution, we employ explicit refinement and derefinement operations. We split a cell into two cells when its mass surpasses twice the target mass resolution and merge it with its neighbours when its mass falls below half of the target mass resolution. In addition, we require that neighbouring cells do not differ by more than a factor of $10$ in volume, otherwise we refine the larger cell. The quasi-Lagrangian approach is intrinsically adaptive to density, resolving higher density regions at better spatial resolution. Moreover, it guarantees that the spatial resolution does not change abruptly at steep density gradients.

Dark matter and stars are treated as collisionless particles and the total gravitational force for all resolution elements is computed using the hierarchical tree method \citep{Barnes1986, Springel2005}. The coupled system of MHD and gravity is then integrated using a second order Leapfrog scheme.

\arepo also includes a hierarchical local timestepping scheme which allows us to integrate every resolution element at the timestep it requires. For a multi-scale problem like galaxy formation this can save significant computational time compared to integrating all resolution elements at the minimum global timestep required for any resolution element. However, it introduces additional challenges for the implementation of source terms as described in Sec.~\ref{sec:timesteps}.
In the following we describe the primary additional source terms that we have added to model the physical processes of radiative cooling, star formation and stellar feedback.

\subsection{Heating \& Cooling}
We model the external UV background as a spatially uniform radiation field that provides a heating term and varies with time. We take its tabulated value from \citet{FaucherGiguere2020} and include it as a heating term in the hydrogen and helium cooling calculation. The UV background sets in at redshift $z\approx9$ and rises until it peaks at $z\approx6-7$. {Atomic hydrogen and helium are self-shielded from the UVB following \citet{Rahmati2013}, who fit the hydrogen shielding column to radiative transfer simulations}. This results in a redshift- and density-dependent self-shielding factor which increases with increasing density from zero to unity. The heating of hydrogen and helium by the UVB is reduced by this factor for every cell at each active timestep. As such, gas cells with densities above $\nH\approx0.01\,\cc$ will have self-shielding factors of $1$, and will not be heated by the UV background. In the idealized simulation we present in this work, we use the $z=0$ values of the UV background. {While heating in the form of photo-ionization and photo-electric heating are important physical processes in the ISM, we have not included these in the current model and discuss the implications of this in Sec.~\ref{sec:discussion}.}

%
%
%

To model the gas cooling, we allow hydrogen and helium to exchange energy via two-body processes including collisional excitation, collisional ionization, recombination, dielectric recombination and free–free emission. Additionally, there is inverse Compton cooling off the cosmic microwave background (CMB). These processes put together
constitute primordial atomic cooling of hydrogen and helium. They
allow gas to cool down to $\sim 10^4$\,K, but become inefficient below this temperature. We
implement them following the chemical network of
\citet{Katz1996}, which is the same as in the {\it Illustris} simulation
described in \citet{Vogelsberger2014}. Once the
primordial gas becomes chemically enriched through the metal ejection
from supernova explosions and AGB mass return, these additional elements can contribute significantly to the
cooling. However, following the injection and ionization states
of each element individually is currently beyond the computational scope of our model. For each gas cell, we trace the abundance of the nine most common elements (H, He, C, N, O, Mg, Si, Mn, Fe) and use these to compute the total metallicity of the cell. Then, we assume collisional ionization
equilibrium and tabulate the heating and cooling rates for processes
not included in the \citeauthor{Katz1996} network. {We note that we use the solar abundances of \citet{Asplund2009} and assume $\Zsun=0.01337$ throughout this work.}

Metal line cooling, low temperature cooling, and molecular cooling
(including H$_2$) are tabulated using the {\small CLOUDY} \citep[][v17.01]{Ferland2017} {equilibrium calculations by \citet{Ploeckinger2020}. We use the publicly available version ``{UVB\_dust1\_CR0\_G0\_shield1}'' that self-consistently models the attenuation by low temperature gas of the UV radiation with a temperature and density-dependent shielding column. This includes the use of {\small CLOUDY}'s ``Orion'' dust model. The input radiation field that \citet{Ploeckinger2020} used during the creation of {\small CLOUDY} is the same as the UV background we use.} During the simulation, the cooling
rate generated by the metal component of a cell is obtained by interpolating the table, which contains the
dimensions of redshift, hydrogen number density, temperature and
metallicity. This is similar to \citeauthor{Vogelsberger2014}, but with the addition of a metallicity dimension to take the metallicity dependency of the cooling rates explicitly into account and an extended range to higher densities. Redshift is tabulated in the range $z=9 \text{ - } 0$, hydrogen number density between $\log (\nH/\cc) = -8$ to $6$, temperature between $\log (T/{\rm K}) = 1 \text{ - } 9 $ and metallicity between $\log(Z/\Zsun) = -4$ to $0.5$. Above the last metallicity bin, the cooling rate is determined by scaling the solar value with the metallicity value in solar units. Below our lowest metallicity we instead use the lowest tabulated value since the cooling rates below $\log(Z/\Zsun) = -4$ are minimally different.
The total cooling rate of a given cell is then computed as
\begin{equation}
  \label{eq:4}
  \Lambda_{\rm cell} = \Lambda_{\rm prim} + \Lambda_{\rm metal} +
  \Lambda_{\rm low\,temp}.
\end{equation}
To extract just the metal cooling contribution we subtract the primordial rates (which are calculated in the \citeauthor{Katz1996} network) from the ones obtained including molecules and metals
\citep[see][for a detailed investigation using this method]{Wiersma2009}. We note that our cooling does not take the relative abundances of
different elements into account. 

Our low temperature cooling is comparable to the fitting formula from \citet{Hopkins2018}, which was obtained by fitting to {\small CLOUDY} modelling including low temperature
metal-line, fine structure and molecular rates, also assuming collisional
ionisation equilibrium \citep[i.e. this is also used in][]{Marinacci2019}.
\begin{figure}
  \includegraphics[width=\linewidth]{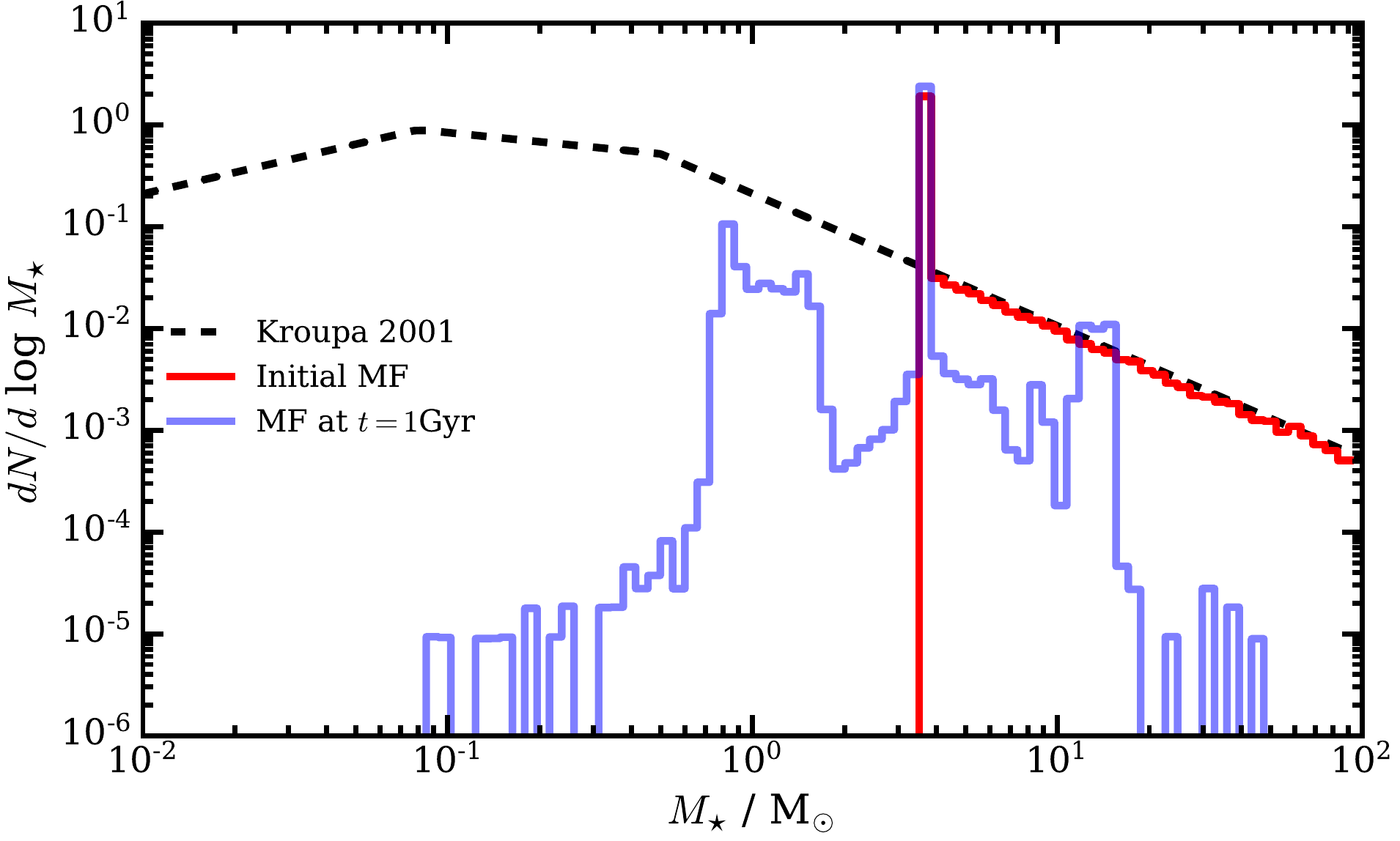}
  \caption[]{Initial stellar mass function and current stellar mass function at later times, as labelled. The IMF below 4 solar masses is \textit{unresolved}, i.e. it is not individually sampled. However, the integrated mass from the unresolved part is distributed into 4 solar mass stars which creates the visible spike at this mass scale. These \textit{unresolved} stars do not lose mass or age, thus this spike also remains at later times.}
  \label{fig:IMF}
\end{figure}
\subsection{Formation of single stars}
\label{sec:SF}
%
Once gas cells are able to cool and contract we expect stars to
form in the densest clumps. 
Star cluster forming clumps have
an average density of $10^3-10^5$\,\cc \citep{Lada2003, Klessen2011}. Thus, we set the onset
of star formation to $\nH>10^3\,\cc$. {A different choice for this parameter does not strongly impact our results (as shown in Sec.~\ref{sec:appendix}). We set the efficiency parameter to $0.02$}. The star formation rate (SFR) $\dot M_{\star,i}$ of a cell $i$ is calculated according to the Schmidt relation
\begin{equation}
  \label{eq:sfr}
  \dot M_{\star,i} =\epsilon_{\rm SF} \frac{M_i}{t_{\rm ff}}, 
\end{equation}
where $\epsilon_{\rm SF}$ is the star formation efficiency and $M_i$ is the mass of the cell. The free fall time of a cell is $t_{\rm ff}=\sqrt{3\pi/(32\, {G} \rho_i)}$.
%
 To have a non-zero SFR, we additionally require the temperature of the gas cell to be  $T<100\,$K.
 
Star formation itself is implemented stochastically, i.e.~the SFR of a cell translates to a probability of creating a star particle in a given timestep. {Each cell can only form a single star per timestep.}
The mass of the potentially created star, $M_{\star \mathrm{, new}}$ is drawn from the (mass-weighted) IMF. The probability $\mathcal{P}$ for a cell $i$ to create a star during a timestep $\mathrm{d}t$ is given by
\begin{equation}
  \label{eq:2}
    \mathcal{P}_i = \frac{M_i}{ M_{\star\mathrm{, new}}}  [1- \exp(-\epsilon_{\rm SF} \frac{\mathrm{d}t}{t_{\rm ff}})]  .
\end{equation}
As can be seen in the red line in Fig.~\ref{fig:IMF}, we truncate our IMF to $M_{\star\mathrm{, min}}${, which is set to $4\,\Msun$ in this paper}. Effectively, we thus follow two types of star particles. \textit{Resolved} stars populate the properly sampled part of the IMF and have masses above $M_{\star\mathrm{, min}}$. These stars follow the stellar evolution pathways described in Section \ref{sec:SN}, meaning they may return mass, metals and energy depending on their initial mass. \textit{Unresolved} star particles all have masses of exactly $M_{\star\mathrm{, min}}$. These star particles do not return mass, metals or energy to the ISM. They merely sample the low mass end of the IMF and provide a gravitational/dynamical contribution. To not overestimate the stellar mass that is to be formed according to the SFR, we guarantee that the total mass of unresolved star particles equals the integrated mass in the IMF that would otherwise be present between 0.01\,\Msun (the lower end of the IMF) and $M_{\star\mathrm{, min}}$. 

Our prescription creates a tail of high density cells whose radii can become much smaller than our gravitational softening length of the gas. Moreover, we do not include the relevant physical processes that dominate at these densities. Thus, we set $\mathcal{P} = 1$ for all cells with $\nH>10^4\,\cc$. This enforced star formation is similar to the treatment by \citet{Lahen2020}. 
As our tests in Fig.~\ref{fig:SFH_tests} show, this choice has a negligible effect on the SFR \citep[see][for a detailed analysis of the effects of increased SF efficiency]{Smith2019}.

In highly resolved simulations where SPH particles or hydrodynamical
cells become less massive than individual stars, the question arises
where to draw mass from to actually form new stars. A common solution to
this issue is to create sink particles that accrete mass for a time until they take on the
characteristics of star particles \citep[e.g.][]{Krumholz2004, Federrath2010, Walch2015b}. This method requires mass to be decoupled from the
hydrodynamic calculation while the sink particle ``grows'' until it can make a
massive star. 
Another solution is to accumulate sufficient star forming material before spawning a massive star. This technique was used in \citet{Hu2017}, but requires instantaneous transfer of mass over the whole galaxy when not enough star forming gas is available locally. This can be avoided by defining a spatially located accretion region \citep[see][]{Hu2019,Hirai2020}.

We opt to create stars instantaneously by defining an accretion
radius for the ``star forming region'' around a star-forming cell \citep[similar to the method in the SIRIUS project presented in][]{Hirai2020}. The
initial guess of the accretion radius is set to encompass all
cells that share an interface with the star forming cell. There are on
average $\sim 20$ of such nearest
neighbours. If the mass enclosed within this radius is
not at least twice the mass of the star to be formed, the radius is
increased until this criterion is met. Once the  radius is found,
the star is formed by drawing an equal fraction $f_{\rm acc}$ of mass from each cell within the accretion radius. This way, the mass of each
cell within the radius is reduced at most by half, and no cells
are fully evacuated due to the formation of the star particle.
The new star inherits the center-of-mass, total momentum, and metal content of the gas that has been removed from cells to form it, which guarantees conservation of mass and metal mass as well as momentum. The stellar metallicity is then calculated as the ratio of the metal mass and the star's mass.

Before creating the stars in a given timestep, the model checks whether any cell is within the accretion region of two or more stars. If this is the case, and if there is not enough mass within the cell to provide the requested mass to each to-be-formed star, formation of the least massive star is cancelled. If needed, this cancellation is repeated until the mass budget is positive. We point out that this a fix for extremely rare circumstances. For example, in our fiducial run this occurred a single time in the 1\,Gyr run time.
\begin{figure}
  \includegraphics[width=\linewidth]{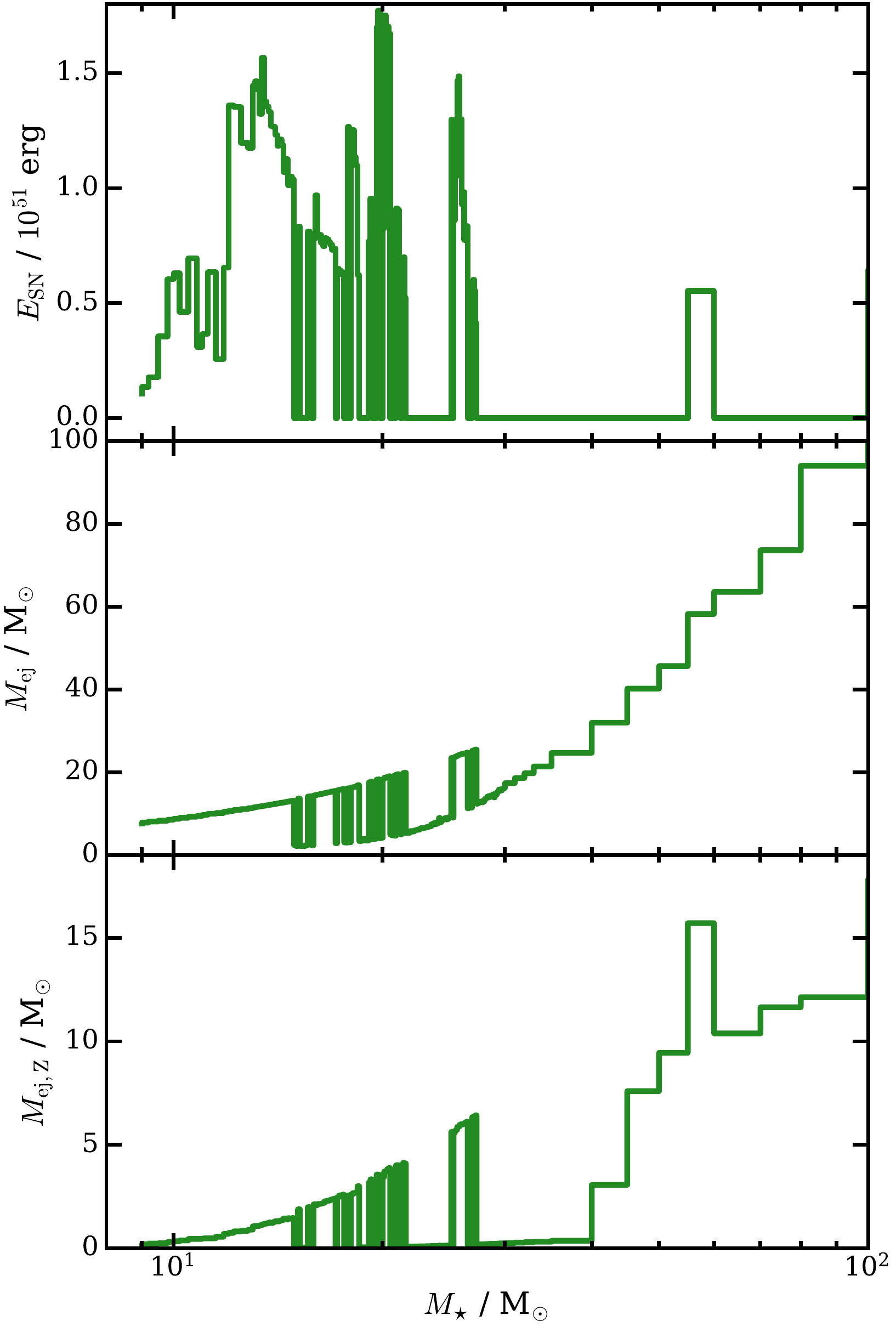}
  \caption[SN energy, mass and metallicity]{Energy, total mass and
    mass in metals ejected per supernova as a function of initial stellar
    mass, from top to bottom. These values include the supernova and wind ejecta presented in \citet{Sukhbold2016}, and are tabulated and
    interpolated during run time of the simulation.}
  \label{fig:SNejecta}
\end{figure}
%
\subsection{Individual supernovae with variable energy}
\label{sec:SN}
As each massive star particle is an individual star, it can produce at most
one supernova within its lifetime. We note that for now we do not take binary stars into consideration. However, in future work we plan to add a prescription for a non-zero binary fraction and account for their modified evolution. Thus, each stellar death induces a discrete explosion event. We ensure that the corresponding blastwave is time-resolved by reducing
the local simulation timestep accordingly (see Section
\ref{sec:timesteps}).

We use the SN explosion models of
\citet{Sukhbold2016} to infer the energy of the supernova (varying from $0
- 1.8\times10^{51}$ erg) from the initial mass of
each star. These models are able to predict a fluctuating energy since they follow the non-monotonic variation in core compactness prior to explosion. They are based on one-dimensional calculations that follow $\sim 2000$ nuclei, assume solar abundance SN progenitor stars, and include neutrino transport. 
When a star reaches the end of its lifetime, we inject the energy, mass and metals according to Fig.~\ref{fig:SNejecta} into the single cell that hosts the star in that particular timestep. 

We choose $M_{\mathrm{SN,min}}=8\Msun$ as our supernova threshold mass. However, according to \citeauthor{Sukhbold2016},
there are some progenitor masses that do not produce
supernovae despite being above $8\Msun$. These stars are assumed to directly collapse to form black holes
instead. Thus, by following this model, we indirectly take this mode of
stellar mass black hole formation into account. The corresponding stars nevertheless eject mass and metals into the ISM. However, we note that our metallicities may be overestimated due to the use of solar abundance SN models (see the Discussion, Section~\ref{sec:discussion}).

As our model does not yet include a stellar wind prescription, we inject both the wind and SN (mass and metal) returns simultaneously. Additionally, we choose to inject a minimum of $10^{49}$\,erg per star above $8\Msun$ at its death to account for some of the (otherwise neglected) wind energy. 
The mean energy injected per star with mass above $M_{\mathrm{SN,min}}$ is $4.4\times10^{50}$\,erg (see Table \ref{tab:energies}). This means that our model injects a factor of $\sim2$ less energy than a model that simply accounts for $10^{51}$\,erg per massive star.
%
%
%
The lifetimes of stars above our supernova threshold are taken from \citet{Portinari1998} and are metallicity dependent.
To test whether our supernovae are Sedov-resolved at our fiducial
resolution, we run separate tests in isolated boxes and check that
a sufficient fraction of the thermal energy we inject is converted
into kinetic energy within the hydrodynamic calculation (see Section \ref{sec:SNboxes}).

\begin{figure*}
  \includegraphics[width=\textwidth]{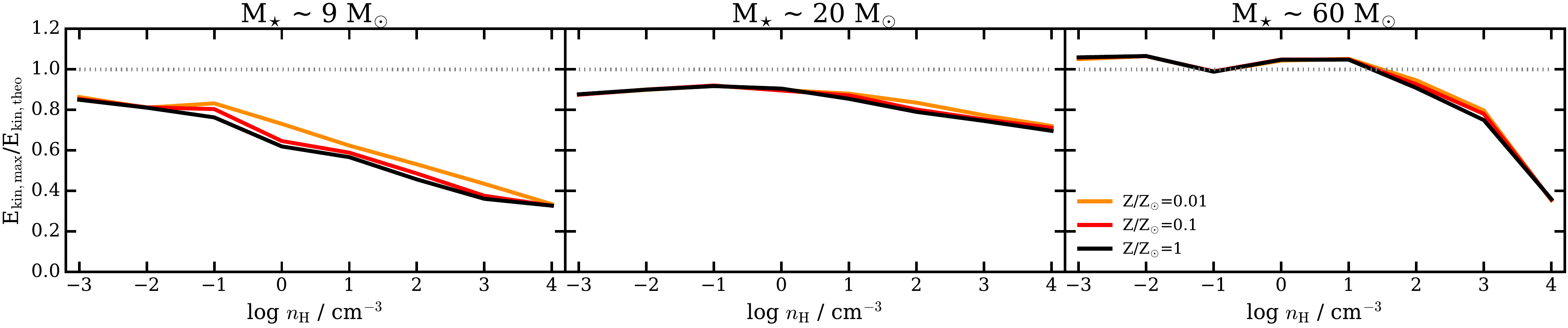}
  \caption[]{Kinetic energy at shell formation time normalized by the expected kinetic energy based on the analytic explosion solution, as a function of the ambient, homogeneous density of the surrounding medium. The panels show the results for a $9\Msun$, $20.5\Msun$ and $59.5\Msun$ star, respectively. The three colors give in each case the results for different metallicities of the ambient medium. All the explosions include energy, mass and metal injection according to our model values shown in Fig.~\ref{fig:SNejecta}. SNe in lower density environments are kinematically more resolved. At almost all tested densities, we reach the expected kinetic energy within a factor of two.}
  \label{fig:SNkinetic}
\end{figure*}

\subsection{Time steps to resolve supernovae}
\label{sec:timesteps}
We integrate the system using individual, variable time-stepping in a
power-of-two hierarchy according to \citet{Springel2010}. The initial
estimate of the hydrodynamical timestep $t_{i}$ necessary for a cell $i$ is determined from the Courant criterion \citep{Courant1967}
according to 
\begin{equation}
  \label{eq:3}
  \Delta t_{i} = C_{\rm CFL} \frac{R_{i}}{c_{i} + |v_{i}|},
\end{equation}
where $C_{\rm CFL}=0.3$ is the Courant factor, $R_{i}=(3V_{i}/4\pi)^{1/3}$, the
spherical radius of the cell, $c_{i}=\sqrt{\gamma P_{i}/\rho_{i}}$ the sound speed
of the cell, and $v_{i}$ the velocity of the cell relative to the
grid. This criterion captures any physical changes happening to the
cell itself.
Additionally, there are a variety of non-local physical processes that may affect the cell. To resolve these, the cell's timestep may need to be reduced below this initial value.
We outline the most
important ones here.
Firstly, the maximum timestep cell $i$ can have while still being aware of the
(potentially supersonic) signals
from other cells can be estimated from the signal velocity $v_{\rm sig}$ of cells in the vicinity \citep[cf.][]{Springel2010}:
\begin{equation}
  \label{eq:sntime}
  v_{\rm sig} = c_i + c_j - {\bf v}_{ij} \frac{{\bf r}_{ij}}{r_{ij}}.
\end{equation}
${\bf v}_{ij}$ is the velocity difference between two cells $i$ and
$j$. The time that it can minimally take for
a signal from cell $j$ to reach cell $i$ is then approximated as $\Delta t_{\rm
  vicinity}=\frac{r_{ij}}{v_{\rm sig}}$. Thus, the timestep of cell $i$ is set to $\min(\Delta t_{i}$, $\Delta t_{\rm
  vicinity})$.

However, this criterion can only be aware of signal
speeds already present within cells. In the case of a supernova
explosion, the energy is injected instantaneously at a given time
independent of the depth of the timestep hierarchy. Consequently, if a
supernova is
injected into a cell that is on a lower timestep than its
neighbors, it is
possible for the supersonic shock to pass the inactive cells before
they are able to react due to their longer timesteps. This
behaviour can be highly insufficient to resolve the Sedov phase of the
supernova shock \citep[see][for a detailed analysis]{Durier2012}.

Fortunately, the timing of the onset of a given
supernova explosion is pre-determined by the initial mass of its
stellar progenitor. Therefore, we can pass the information of an
imminent supernova to the timestepping routines \textit{before} the supernova is injected.
Consequently, the cells are granted a lead time to adjust their timesteps suitably. This is accomplished by tracking all stars that
will explode within the next global timestep.
For a cell $i$ that hosts one such star, the sound speed that it would
have if the supernova were injected is
calculated according to
\begin{equation}
  \label{eq:6}
  c_{\rm SN} = \sqrt{ \gamma(\gamma-1) \Big(u_i + \frac{E_{\rm SN}}{M_i}\Big)},
\end{equation}
where $\gamma$ is the adiabatic index, $u_i$ is the internal energy
per unit mass, $M_i$ the mass of the cell and $E_{\rm SN}$ is the energy of the SN
(which can vary, see Section \ref{sec:SN}). To obtain the desired
result of making sure the cells in the vicinity of an imminent SN have
low enough timesteps to guarantee a time-resolved blastwave, it would be sufficient
to set $c_{\rm SN}$ in eq. (\ref{eq:sntime}) starting at the last global timestep (defined as a
timestep where all cells are active) prior to
injection. This, however, slows down the calculation considerably and
partly unnecessarily. The timestep hierarchy in our isolated runs is usually 5-6 levels deep, meaning that the smallest timestep can be $2^6=64$ times smaller than the global timestep.
So, if for example the SN event happens towards the end of a global
timestep, the region around the ``to-be'' SN will likely be needlessly on the lowest
timestep  for the majority of the time. 

Instead, we can steadily shorten the
timesteps toward the required timestep $\Delta t_{\rm SN} = R_i / c_{\rm SN}$. 
Hence, we calculate a timestep $\Delta t_{\rm max}$ for each SN host cell 
\begin{equation}
  \label{eq:5}
  \Delta t_{\rm max} = f_{\rm SN} (t_{\rm SN} - t_{\rm now}) ,
\end{equation}
where $t_{\rm SN}$ is the time
of the supernova, $t_{\rm now}$ is the time of the simulation and
$f_{\rm SN} = 0.9$ is an ad hoc factor, the choice of which does not affect the results presented here.
This requires that the cell will be active at least one more time before the SN
event. This calculation is repeated again at the next active
time. Thus, the timestep is iteratively reduced until $c_{\rm SN} >
R_i/\Delta t_{\rm max}$.
This ensures that the cells
surrounding the imminent supernova will reduce their timesteps
according to Eq.~(\ref{eq:sntime}). When the supernova is finally
injected, the cells within reach of the signal speed of the shock will
be prepared.

We note that the star can move from one host cell to another before
the SN is injected. In this case, the new cell's timestep is reduced
instead. We do not expect this to affect the outcome, since the new
cell is likely a neighbor of the former host. As such, its
timestep has already been reduced to a similar degree.

\subsection{AGB mass return}
\label{sec:agb}
%
At present the only form of non-SN stellar feedback in our model  is
the mass and metal return from asymptotic giant branch (AGB) stars (although we do account for some stellar wind energy, see Section~\ref{sec:SN}). For AGB stars, we
tabulate the mass return for each element as a function of initial stellar mass and stellar metallicity, following the models of \citet{Karakas2018}. For simplicity, the mass return is injected all at once at the end of life of the star. The lifetimes of AGB stars are also determined from \citeauthor{Karakas2018} and are initial mass and metallicity dependent. The mass of the star particle is reduced by
this amount and the mass of the host gas cell is increased. Additionally, the mass for each element is increased by the metal mass created. The
hydrodynamic distribution of the mass is then handled by \arepo's
cell refinement.
%
%
%
%

%
\begin{table}
  \centering
    \caption{Progenitor mass, supernova energy, mass return (from winds and SN) and metal mass return for the three SN tests shown in Fig. \ref{fig:SNkinetic}.}
  \begin{tabular}{c|c|c|c}
\hline\hline
    $M_{\mathrm{init}} /\Msun$& $E_{\mathrm{SN}}/ 10^{51}$\,erg & $M_{\mathrm{ej}} /\Msun$&$M_{\mathrm{ej, Z}} /\Msun$\\
\hline
9&0.1066&7.64&0.17\\
20.5&1.6714&19.03&3.81\\
59.5&0.4972&56.97&4.87\\
    \hline\hline
  \end{tabular}
  \label{tab:SNTests}
\end{table}
\section{Tests for individual supernovae}
\label{sec:SNboxes}
%
A problem often faced by cosmological
galaxy formation simulations that include SN feedback is that at the typical
spatial and temporal resolution that can be afforded the thermal energy injected by SNe is quickly
radiated away in cooling before hydrodynamics can transform a
fraction of this energy into momentum. When this happens, feedback is
numerically inefficient at heating and removing dense gas and preventing
it from forming stars. In the literature, there are a variety of prescriptions to ameliorate this issue, i.e.~injecting the energy as momentum \citep[e.g.][]{Navarro1993, Springel2003, DallaVecchia2008}, preventing gas from cooling for a time after the SN is injected \citep[e.g.][]{Stinson2006}, storing up energy until there is enough for the desired effect \citep{Schaye2015} or adding additional momentum based on an estimate of how much energy will be numerically lost \citep{Hopkins2018}. \citet{Naab2017} provide an excellent review of the various prescriptions and their merits.

Instead of following any one of these solutions, we {set the mass resolution high enough to self-consistently produce most of the momentum from a thermally injected SN}. The theoretical ``similarity solution'' for a blastwave predicts an
early energy conserving phase known as the Sedov-Taylor phase during
which the total energy remains constant while the shock front moves outward. This phase ends when radiative cooling in the dense shell
formed by the shock becomes dominant and energy is no longer conserved. At this time, the total energy is predicted to be distributed in
thermal and kinetic energy following the ratio 72:28 \citep{Chevalier1976, Ostriker1988}. If indeed 28\%
of the initial energy is converted to kinetic energy numerically, the SN is fully Sedov-resolved.

We test to what extent the SNe in our simulation are Sedov-resolved \citep{Sedov1959} by
simulating individual SN explosions in isolated, uniform-density boxes at the
fiducial resolution of our main simulations {($4\,\Msun$ target gas mass)}. This is similar to the work by
\citet{Kim2015}, \citet{Hu2016}, \citet{Haid2016} and \citet{Steinwandel2020}. To this end, we set up a
grid of
initial conditions with varying values of ambient density, ambient
metallicity and stellar progenitor mass. The bins are log($\nH /\cc) = -3 \text{ - } 4$ in steps of $1$, log($Z/\Zsun$) = \{$-2$, $-1$, $0$\}. The stellar masses are tested for three values of $9\Msun$, $20\Msun$ and $60\Msun$. The progenitor mass then sets
the SN
energy with matching amounts of mass and metal ejecta, see Table \ref{tab:SNTests}. We set the ambient thermal energy of the gas following
the equilibrium curve in our phase diagrams. 

For each SN test box, we follow the total energy within the remnant,
and how much kinetic energy is induced by the blastwave. Fig.~\ref{fig:SNkinetic} shows the maximum kinetic energy relative to the kinetic energy predicted by the analytic solution. This is $28\%$ of the energy of the supernova. We show this energy ratio as a function of ambient density.

As can be seen in Fig.~\ref{fig:SNkinetic}, the numerical results move towards the analytical solution for decreasing ambient densities. Since there is less mass to sweep up, the blastwave reaches the shell formation time later \citep[see eq.~7 in][]{Kim2015}, and the shell is physically larger, giving the code time to resolve the processes better. {As we will see in Fig.~\ref{fig:SN_density_ideal}, the majority of our SNe explode in low density regions. Thus, we judge $>90\%$ of the SNe to be well resolved.} Interestingly, high initial energies are also better resolved. This can be explained by the higher energy contrast between the initial injection energy and the background temperature. There is almost no dependence on metallicity, although lower metallicities give marginally more accurate results due to the cooling rates being lower.
{We have also tested the hot gas fractions produced by our SN and compared these to theoretical expectations. They are well matched within a factor of two.}

We note that tests with a multi-cell injection scheme yield a lower amount of kinetic energy than what is shown in Fig.~\ref{fig:SNkinetic}. 
By injecting the energy into multiple cells, we increase the initial radius. In the case of high densities, the radius thus approaches the Sedov radius itself. This in turn renders the blastwave less well resolved.
\begin{figure}
  \includegraphics[width=\linewidth]{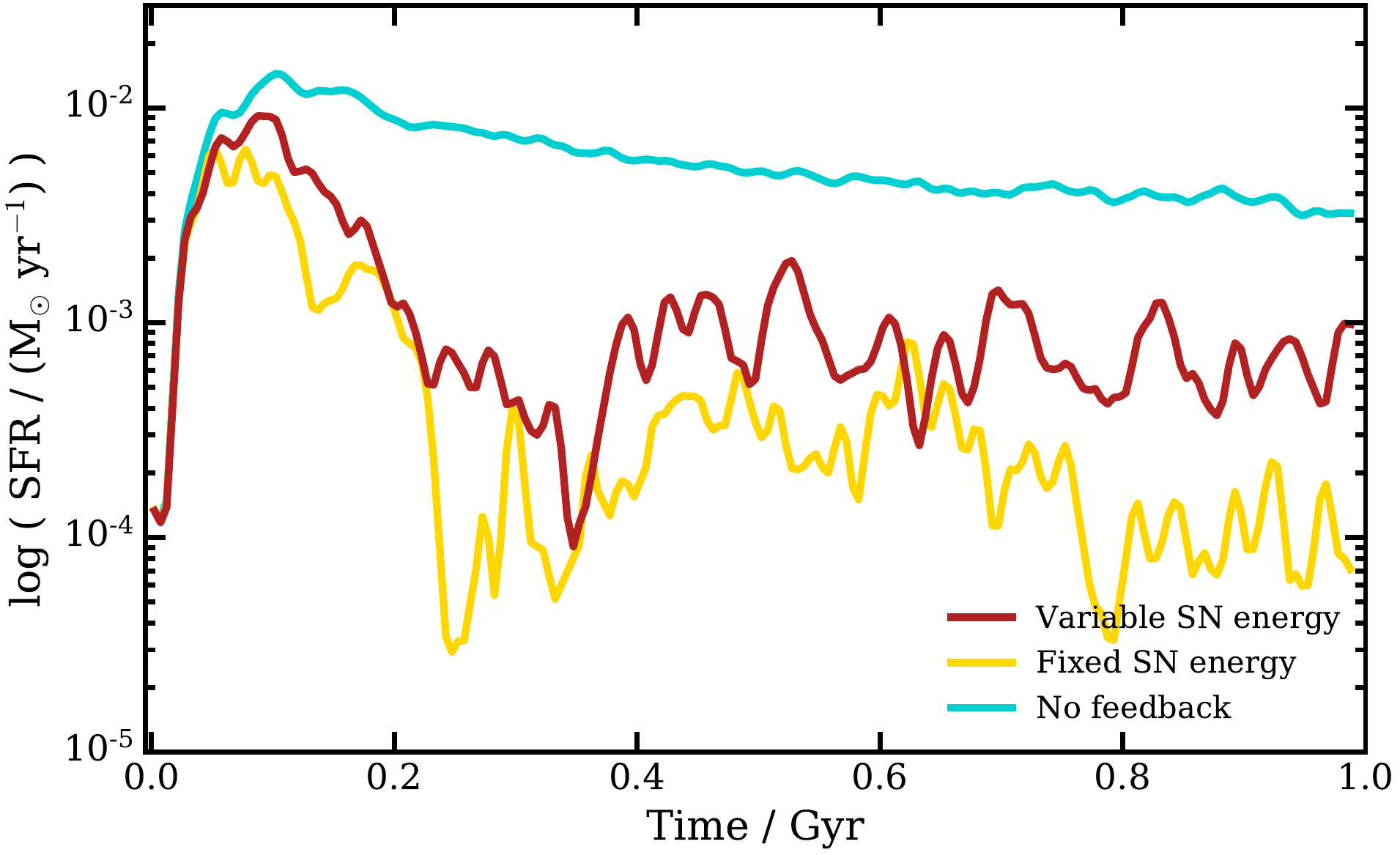}
  \caption[]{Star formation histories of our various isolated simulation runs. Models with feedback begin reducing their SFR after the first burst, and retain a small SFR of $\sim10^{-3}\Msun {\rm yr}^{-1}$ after 1\,Gyr. Models without feedback yield a much higher SFR of $\sim10^{2}\Msun {\rm yr}^{-1}$ for the majority of the simulation time.}
  \label{fig:SFH_ideal}
\end{figure}
\begin{table*}
  \centering
    \caption{Total injected energy, mean energy injected per SN, mean energy injected per massive star ($>8\,\Msun$), mean energy injected per total stellar mass and number of SNe within the 1\,Gyr run time of the simulation.}
  \begin{tabular}{l|c|c|c|c|c}
\hline\hline
    Name&$E_{\mathrm{inj}}$& $E_{\mathrm{inj}} / N_{\rm SN}$&$E_{\mathrm{inj}} / N_{\star}(\rm >8\Msun)$&$E_{\mathrm{inj}} / M_{\star}$&$N_{\rm SN}$\\
    &[$10^{54}$\,erg]&[$10^{51}$\,erg]&[$10^{51}$\,erg]&[$10^{49}$\,erg/\Msun]&[$10^3$]\\
\hline
\varSN&$6.75$&$0.57$&$0.44$&0.46&11.7\\
\fixSN&7.82&1.0&1.0&1.03&7.8\\
    \hline\hline
  \end{tabular}
  \label{tab:energies}
\end{table*}
\begin{figure}
  \includegraphics[width=\linewidth]{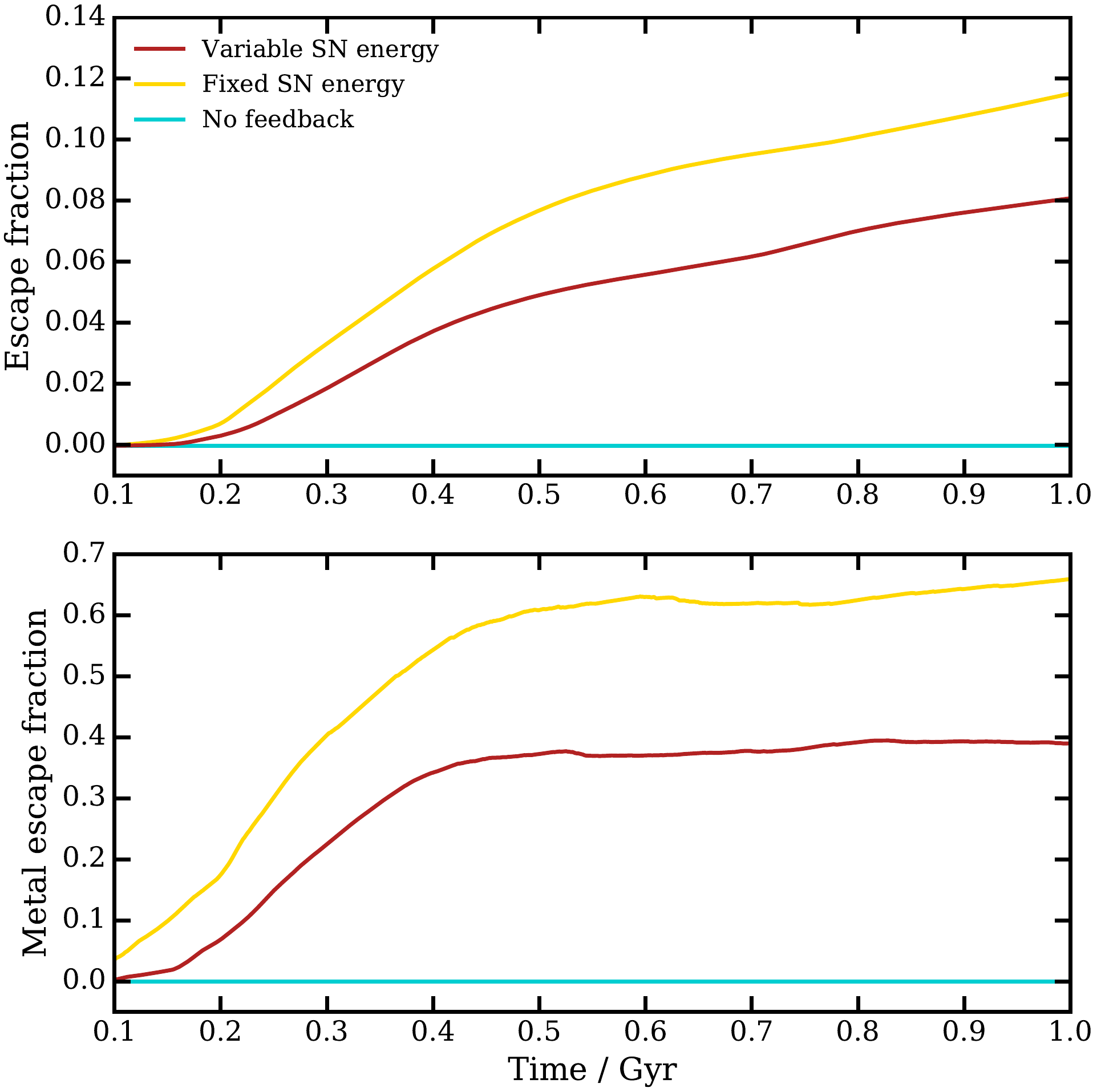}
  \caption[]{Fraction of gas that escapes the virial radius over time (\textit{upper panel}) and fraction of metals produced that escape (\textit{lower panel}) in our various isolated runs with SNe. Runs without SNe have zero escaped material. \varSN loses less mass and less metals than the fixed energy models.}
  \label{fig:escape_frac_ideal}
  \end{figure}
  
  \begin{figure}
  \includegraphics[width=\linewidth]{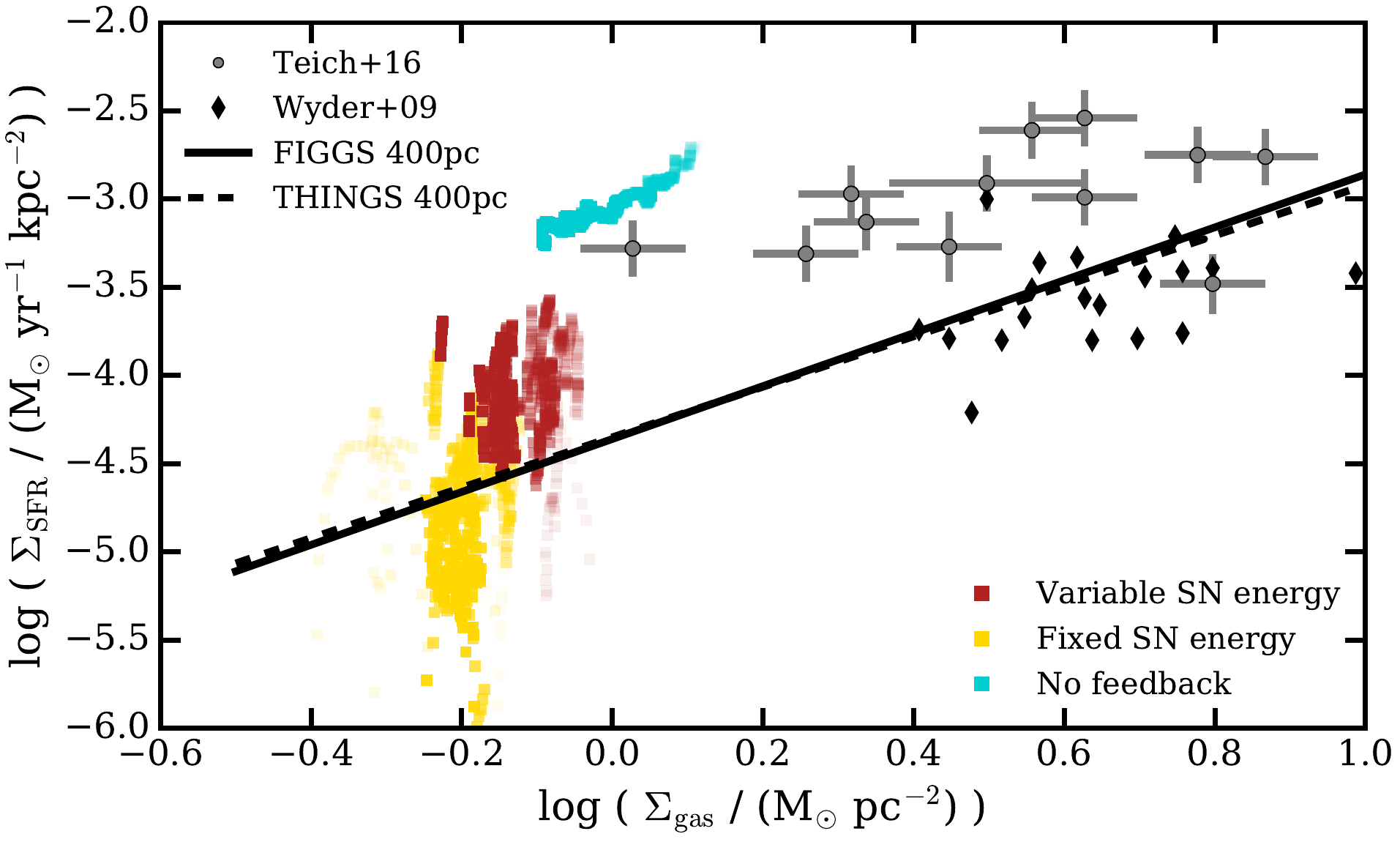}
  \caption[]{Kennicutt-Schmidt relation of our various isolated runs. Squares symbols are colored according to the run, with low alpha values (high transparency) showing earlier times. Each data point is averaged over $10$\,Myr and over a cylindrical disk with a radius set by the maximum radius of stars at that time, and having a height of $200$\,pc. The grey error bars give the observed low mass galaxies from the SHIELD sample \citep{Teich2016}. Black diamonds show the observations presented in \citet{Wyder2009}. The solid black line represents the best fit from the FIGGS sample of dwarf irregular galaxies \citep{Roychowdhury2014}, while the dashed line is the best fit from the THINGS sample \citep{Walter2008}. 
  }
  \label{fig:KS_relation_ideal}
\end{figure}
%
\section{Results for an isolated dwarf galaxy}
\label{sec:isolated}
To validate our new model, we present results from an idealized dwarf galaxy. For easy comparison, we choose identical initial conditions to the ones used in \citetalias{Hu2017} {and run the simulation at a mass resolution of $4\,\Msun$. The initial conditions} consist of a live dark matter halo with a Hernquist profile. The total halo mass is $M_{\rm halo}\approx2\times10^{10}\Msun$. Background stars and gas are placed within the halo following an exponential profile with a scale length of $0.73$\,kpc. The mass of the stellar disk is $2\times10^7\,\Msun$, and that of the gas disk  $4\times10^7\,\Msun$. {The scale height of the stellar disk is 0.35\,kpc. Initially, there is no gas beyond the disk.} The initial setup was run for $200$\,Myr including supernova explosions to create quasi-stationary turbulence within the gas. {This was done by \citetalias{Hu2017} to create more realistic initial structures and add pressure support to prevent early collapse. As we will see, these initial conditions do not produce the same effect in our model.} We note that all following analysis includes only stars that form during the actual run time of our simulation, not the ones from the initial  $200$\,Myr equilibration period or the initial conditions. Differently than \citetalias{Hu2017}, we choose to give the gas an initial metallicity of $0.01\,Z$/\Zsun (instead of 0.1). {The value expected from observations of Local Group dwarfs \citep[i.e.][]{Kirby2013} at this stellar mass is 2-3\%}. To transform the SPH initial conditions into ones usable for \arepo, we add a background grid of gas cells within a cubic box with a boxsize of $12\,$Mpc.
During run time, \arepo refines and de-refines cells as needed to adhere to our mass refinement criterion \citep[see][]{Springel2010}. Within \Rvir there are approximately 7.4 million gas cells and 2.8 million dark matter particles.
%
\subsection{Model variations}
We compare {three versions of our model}:
\begin{enumerate}
    \item 
 \varSN: the fiducial set up as described in Section~\ref{sec:model}. 
\item
\fixSN: for this run we allow every star with initial mass above $M_{\mathrm{SN,min}}$ to be a SN with $10^{51}$\,erg. The mass and metal return remains as shown in Fig.~\ref{fig:SNejecta}. Cumulatively, this model returns {a factor of $\sim 2.2$ more energy per stellar mass} than \varSN (see Tab.~\ref{tab:energies}). To a large degree, this is due to approximately $25\%$ of the high mass stars in \varSN not forming SN but collapsing directly to black holes while ejecting mass and metals. This model is most closely comparable to the common assumptions made in cosmological simulations that sample stars via single stellar population (SSP) particles and integrate the IMF to compute the returned SN energy.
\item
\nofb: in this simulation, we switch off all mass, metal and energy returns from core collapse SN. Any metallicity increase in this run is due to AGB star returns alone.
\end{enumerate}
\begin{figure}
  \includegraphics[width=\linewidth]{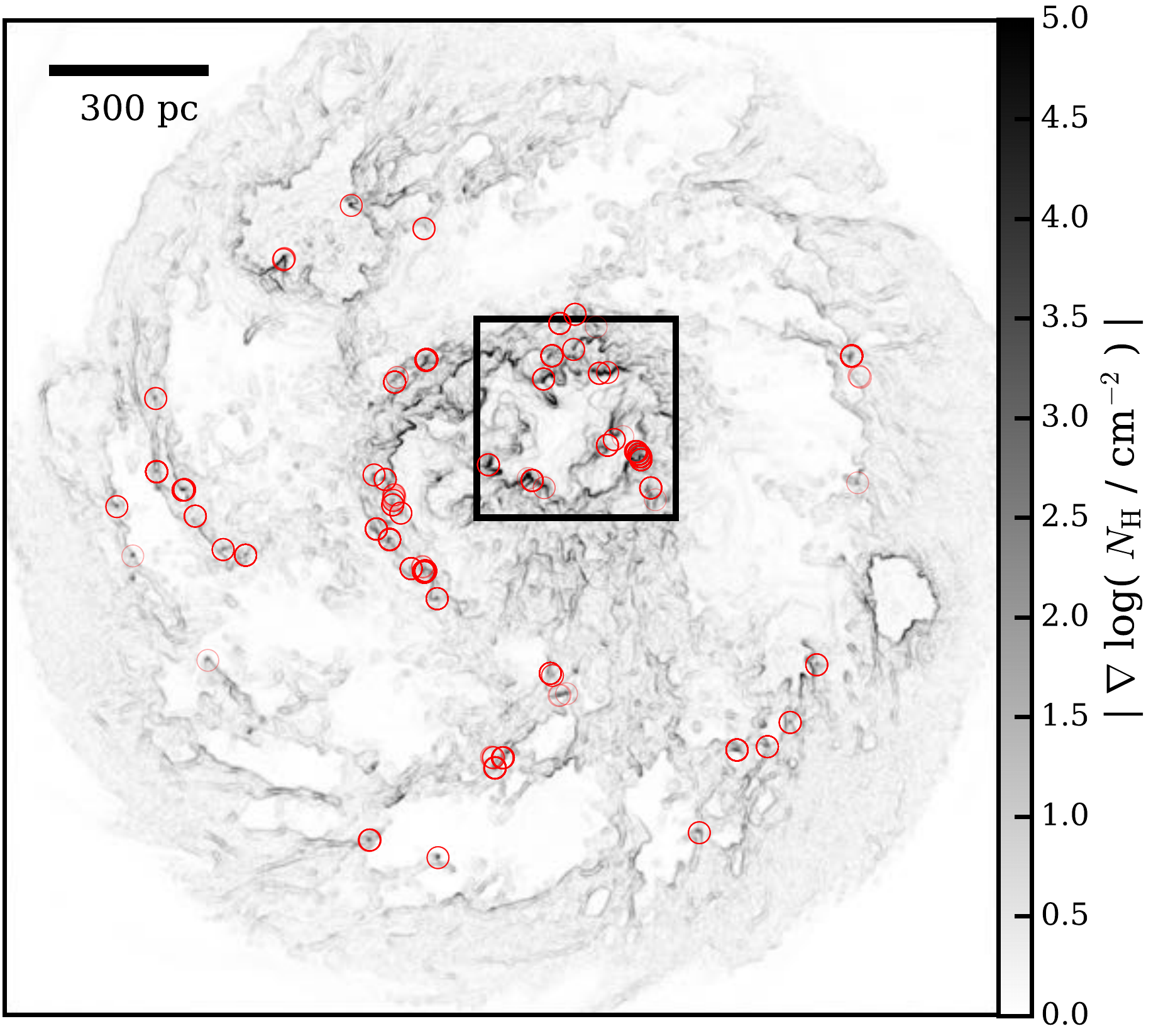}
  \includegraphics[width=\linewidth]{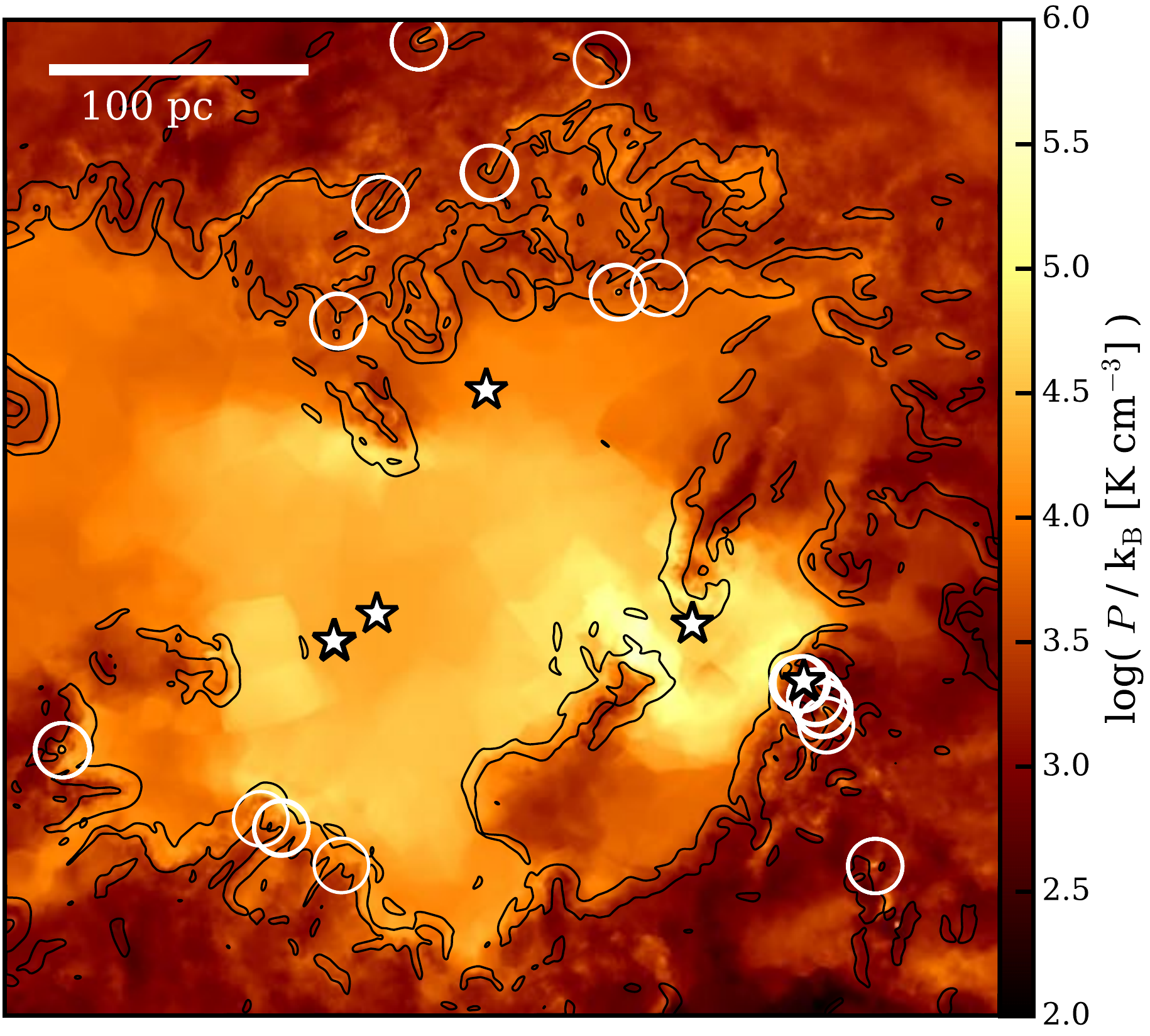}
  \caption[]{\textit{Upper panel:} Column density gradient in grey scale. Red circles indicate new SF sites within the last 1\,Myr. The black square indicates the position of the zoom in the panel below. \textit{Lower panel:} The color map shows the pressure, while the black contours indicate the peak density gradient. White circles are  sites of new star formation within the last 1\,Myr. White star symbols indicate locations of SNe between  $1 < t_{SN}/\mathrm{Myr} < 1.2$. The displayed simulation time is $t=50\,{\rm Myr}$.}
  \label{fig:triggeredSF}
\end{figure}
\begin{figure*}
  \includegraphics[width=\textwidth]{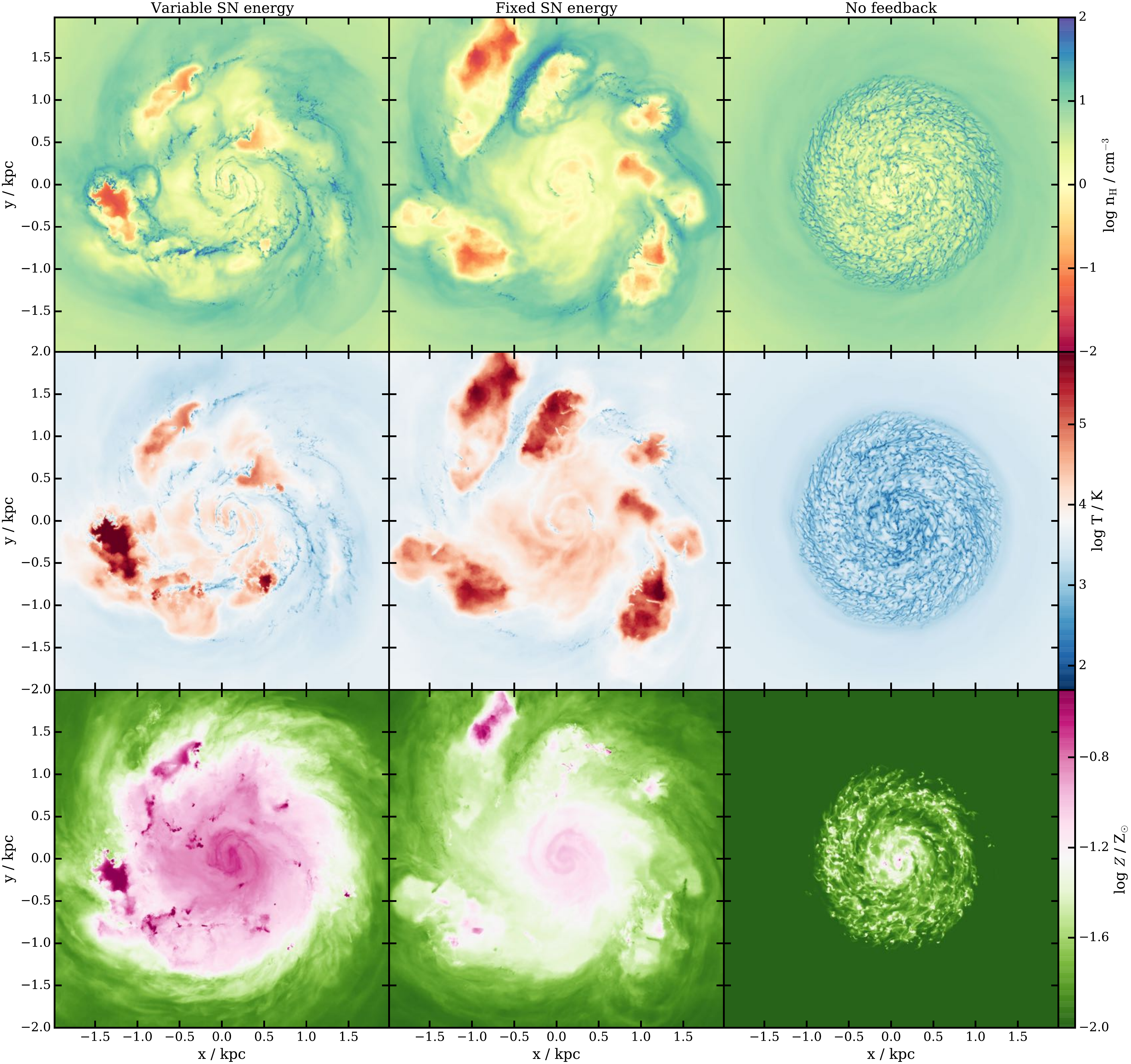}
  \caption[]{Face on density-weighted projections through the gas of our various isolated runs. \textit{Upper row}: hydrogen number density, \textit{middle row}: temperature, \textit{lower row}: metallicity. As labelled above the panels, the different columns correspond to the different runs, from left to right: \varSN, \fixSN, and \nofb, respectively. The simulation time is $500\,$Myr.}
  \label{fig:proj_ideal}
\end{figure*}
\begin{figure*}
  \includegraphics[width=\textwidth]{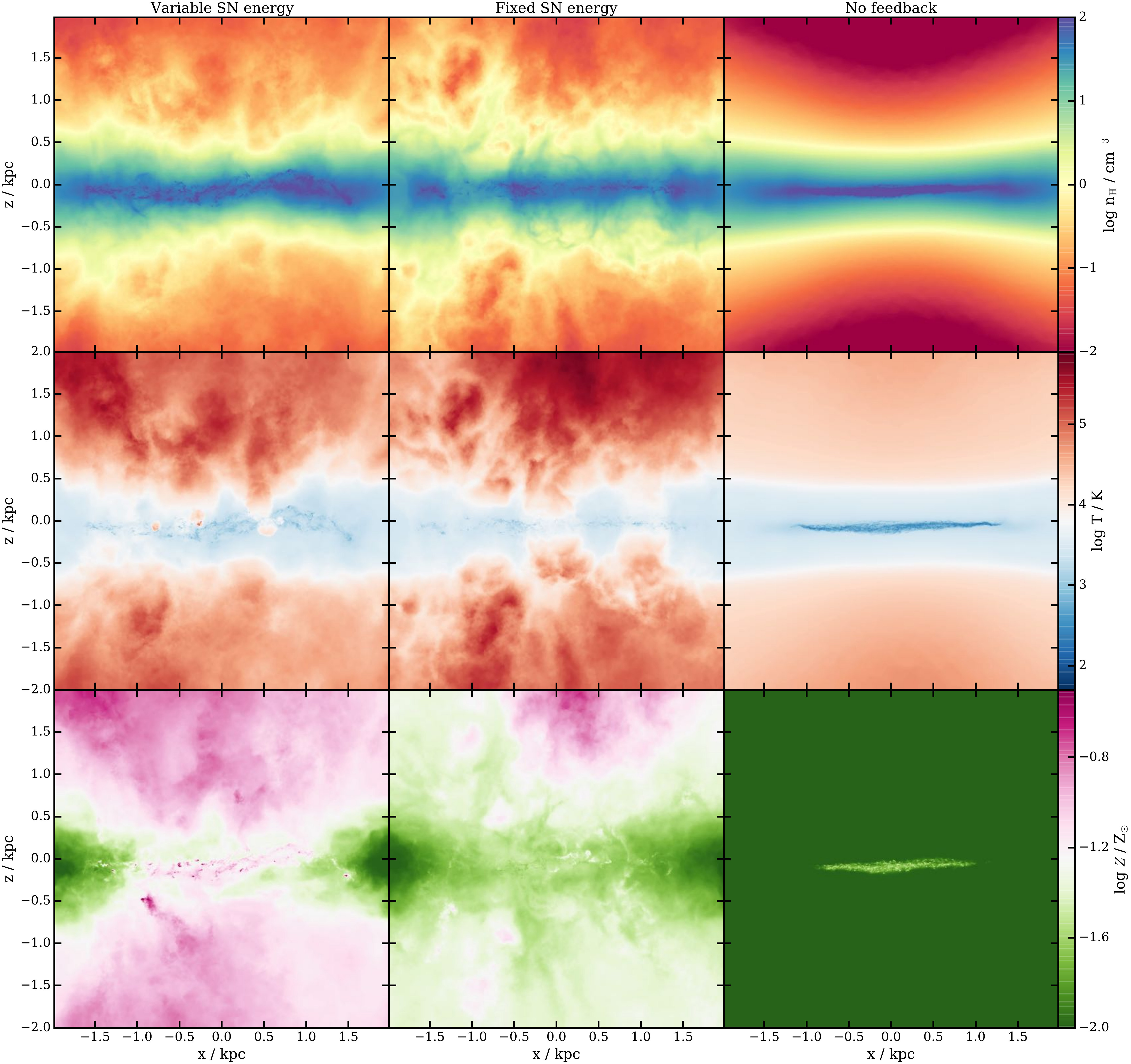}
  \caption[]{Same as Fig.~\ref{fig:proj_ideal} but edge on projections.}
  \label{fig:proj_edge}
\end{figure*}
\begin{figure*}
  \includegraphics[width=\textwidth]{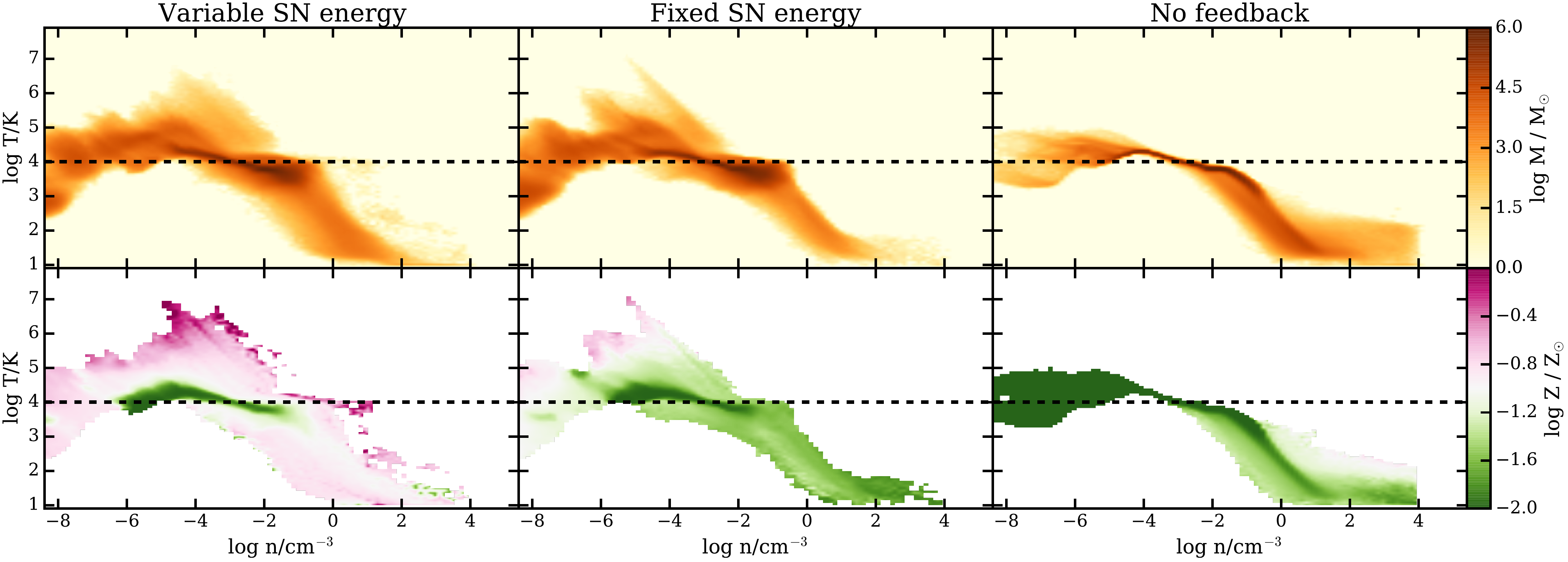}
  \caption[]{Mass (\textit{upper panels}) and metallicity-weighted (\textit{lower panels}) phase diagrams of our various isolated runs. Models with feedback exhibit a clear hot phase at lower densities (top left quadrant) that models without feedback lack. Metals are injected with the SNe, so the hot phase is also metal enriched relative to the disk gas which mainly sits a $10^4$\,K.}
  \label{fig:phase_ideal}
\end{figure*}
%
\subsection{Star formation history and the Kennicutt-Schmidt relation}
%
We begin by analyzing the star formation history (Fig.~\ref{fig:SFH_ideal}) of these five models. Since this is an idealized set up, the SFR is highly dependent on the density distribution in the initial conditions. The SFR shows a strong increase that is similar across all runs during the first 80-100\,Myr. After this initial time, we see the effects of the differing models. \varSN increases until the energetic feedback from SNe begins to heat and blow out gas, decreasing the SFR. After about 400\,Myr a bursty but overall steady SFR emerges at around $10^{-3}\,\Msun \yr^{-1}$. This is also the case for the \fixSN run, albeit the process begins earlier (so the peak is not as high) and the steady value is lower, since there is almost twice the amount of energy per SN (see Table \ref{tab:energies}). {To investigate the driver of these differences, in Fig.~\ref{fig:escape_frac_ideal} we show the mass and metal escape fractions beyond the virial radius. It is clear that \fixSN loses a much larger fraction of its gas mass and metals. Two thirds of the produced metals end up beyond the virial radius and do not contribute to cooling and SF. This lack is the main driver of the lower SFR in this model.} The SFR of \nofb does not { begin decreasing until a few Myr after \varSN. When it does,} the curve flattens out and then begins to drop slowly due to the lack of fuel for new stars. 

We investigate whether the star formation (SF) may be triggered by the SNe that send shocks through the surrounding disk, compressing the gas. The top panel of Fig.~\ref{fig:triggeredSF} shows a face-on projection of the disk where the newly formed stars (formed within the last 1\,Myr) are represented as red circles. They primarily sit on or near where the density gradient (in grey) is strongest. The lower panel shows a zoomed in representation of the area within the black square. Here, the colormap shows the pressure, while the peaks in the density gradient are indicated by black contours. Recent SF is now shown as white circles, while recent SN sites are indicated by white star symbols. It seems evident that the SNe carve out pressurized bubbles that expand. Where these bubbles run into material, they compress the gas, seen as the higher pressure at the edges of the bubbles. This is where new stars begin forming after the SN occurs. We even see a SN within the star forming sites that is a short-lived, massive star that has already exploded.

Relating the available cold gas to the mass in stars formed in these models and comparing this to the observed Kennicutt-Schmidt relation is an informative way to understand whether our star formation prescription performs well. Fig.~\ref{fig:KS_relation_ideal} shows our models relative to observed galaxies. Since our initial conditions represent a dwarf galaxy, it is not surprising that our points end up at the very lower end of the Kennicutt-Schmidt relation. So much so that any comparison is uncertain, since this region of the relation is not well constrained. In general, is is however encouraging to see that all models with feedback lie close the extrapolation (solid and dashed lines) of the relation. In contrast, \nofb clearly has a too high SF for the available gas density. {\varSN exhibits a SFR consistently above the lines.} This is not surprising, since our model at present does not include any other form of energetic feedback except for core collapse SN. Previous simulations have shown that the inclusion of radiative processes, magnetic fields or stellar winds all work in the direction of reducing the SFR \citep[e.g.][]{Gatto2017, Peters2017, Girichidis2018b}. {Interestingly, the SFR of \fixSN may be slightly lower than expectations, showing that this feedback prescription may be too strong. This model expels a larger fraction of the gas and metals (see Fig.~\ref{fig:escape_frac_ideal})}. Note also that a truly fair comparison to observations requires detailed mocking that we postpone to a later paper.
%
%
%
\subsection{Projections and phase diagrams}
To gain a qualitative understanding of the differences between these models, we show face-on projections (projecting down the $z$-axis, assuming the disk lies in the $xy$-plane) of the gas in Fig.~\ref{fig:proj_ideal}. The top row shows the hydrogen number density in a $6\times6\times3$\,kpc box. The central row shows the mean temperature along the line-of-sight ($z$-axis), and the bottom row shows the mean metallicity. {\varSN has more cold and dense cloud structure in the inner 1\,kpc radius and a higher metallicity in this area. Due to the stronger feedback in \fixSN, the region affected by the SN is less dense and lacking in cold clouds. The metallicity is much lower in the central regions. In Fig.~\ref{fig:proj_edge} we see the same but projected down the $y$-axis showing the edge-on view of the disk. Especially in the temperature projection it becomes apparent that the fixed energy model exhibits wider chimneys that expel more hot gas. The stronger outflows also significantly reduce the metallicity, not only in the disk but even beyond.}
The model that is most visibly different is \nofb which exhibits no low density or high temperature gas. \nofb also has almost no metallicity above the background metallicity of $10^{-2}\,\Zsun$. The small enhancement is due to the AGB metal returns that is almost irrelevant on timescales of the runtime of our simulations.

Another useful diagnostic is the phase diagram of gas. This is the two-dimensional distribution in the density--temperature plane. We show the mass (top row) and metallicity-weighted (bottom row) phase diagrams for all five models in Fig. \ref{fig:phase_ideal}. The dashed line in each panel shows $T=10^4$\,K, which is a common temperature cut-off in cosmological simulations and represents the approximate threshold between atomic and molecular cooling processes. 
The upper left quadrant contains hot and low density gas that is heated by SN and then expands to lower densities. Since the SNe inject large amounts of metals, the metallicity in this gas phase is extremely high until it is mixed with the surrounding medium. Clearly, \nofb does not exhibit any gas in this region. Without SNe, the gas follows the equilibrium curve. Cells in the lower left quadrant are cold and diffuse. These are cells beyond the virial radius that have expanded adiabatically into the background volume of the simulation. Gas cells populating the lower right quadrant are cool and dense. These cells follow the equilibrium cooling curve to lower temperatures and higher densities. The ``width'' of the phase structure at temperatures $T=10^2-10^4$\,K and densities between $n_{\rm H} = 10^{-2}-10^2$\,\cc is due to the metallicity dependent cooling. Higher metallicity gas cools at a faster rate, thus dropping its temperature while the density is still relatively low. In other words, high metallicity gas cools first and then condenses, while low metallicity gas condenses and then cools. 

This behaviour is best visible in \nofb because there are no SNe to mix the metals. Here we see a strong metallicity gradient in the $<10^4\,$K gas. In the \fixSN model, the cold and dense gas is less metal rich. This is due to the stronger outflows caused by the enhanced SN energy that eject a larger fraction of the metals (see Fig.~\ref{fig:escape_frac_ideal}).
Another aspect to note is the metallicity distribution in the $>10^4\,$K gas in the runs with feedback. Here, the gas transitions from being metal-enriched (pink) to having lower metallicity (green) while it cools from $10^8$\,K down to $10^5$\,K. This gives an indication of the strength of metal mixing. \fixSN mixes more strongly.
\begin{figure}
  \includegraphics[width=\linewidth]{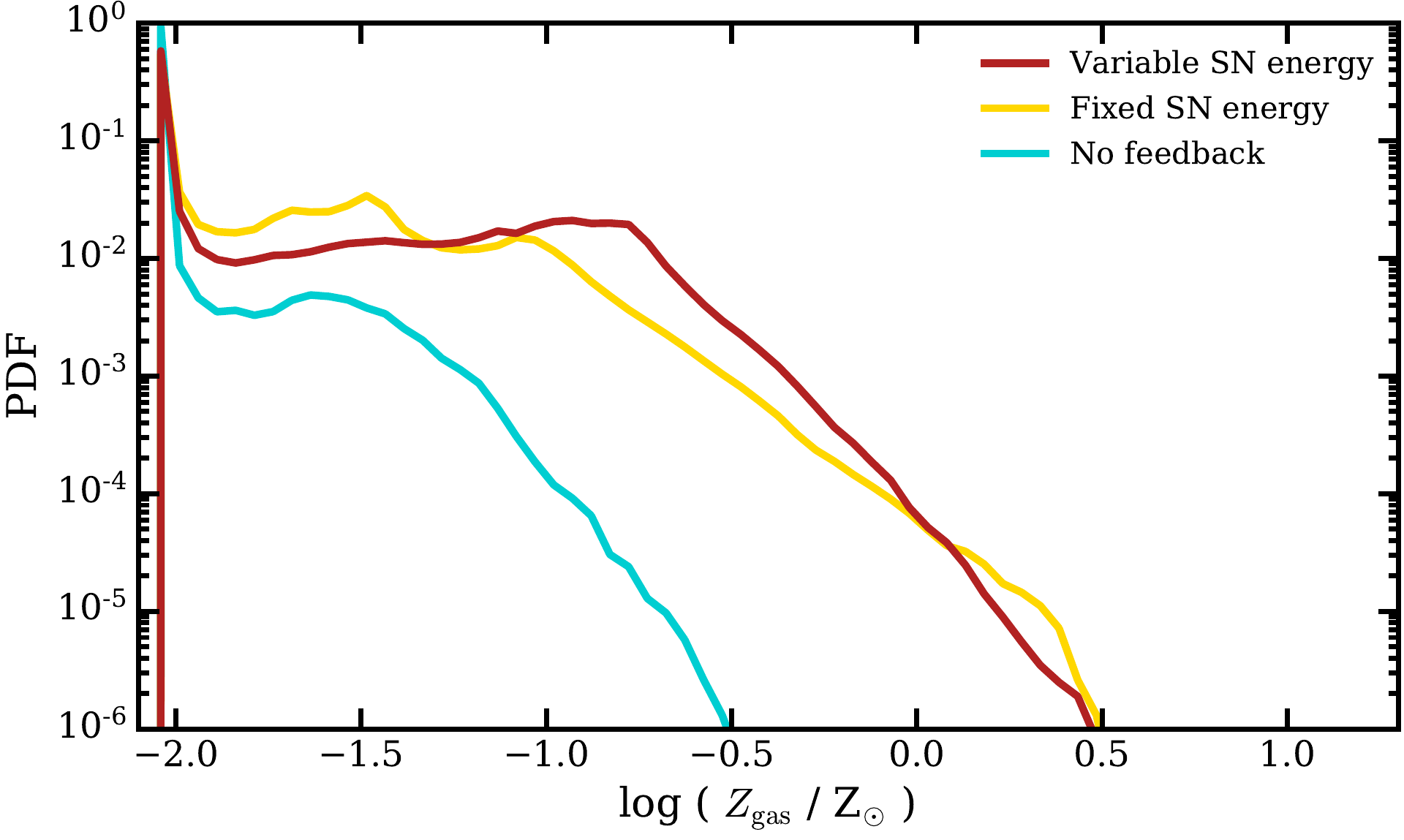}
  \includegraphics[width=\linewidth]{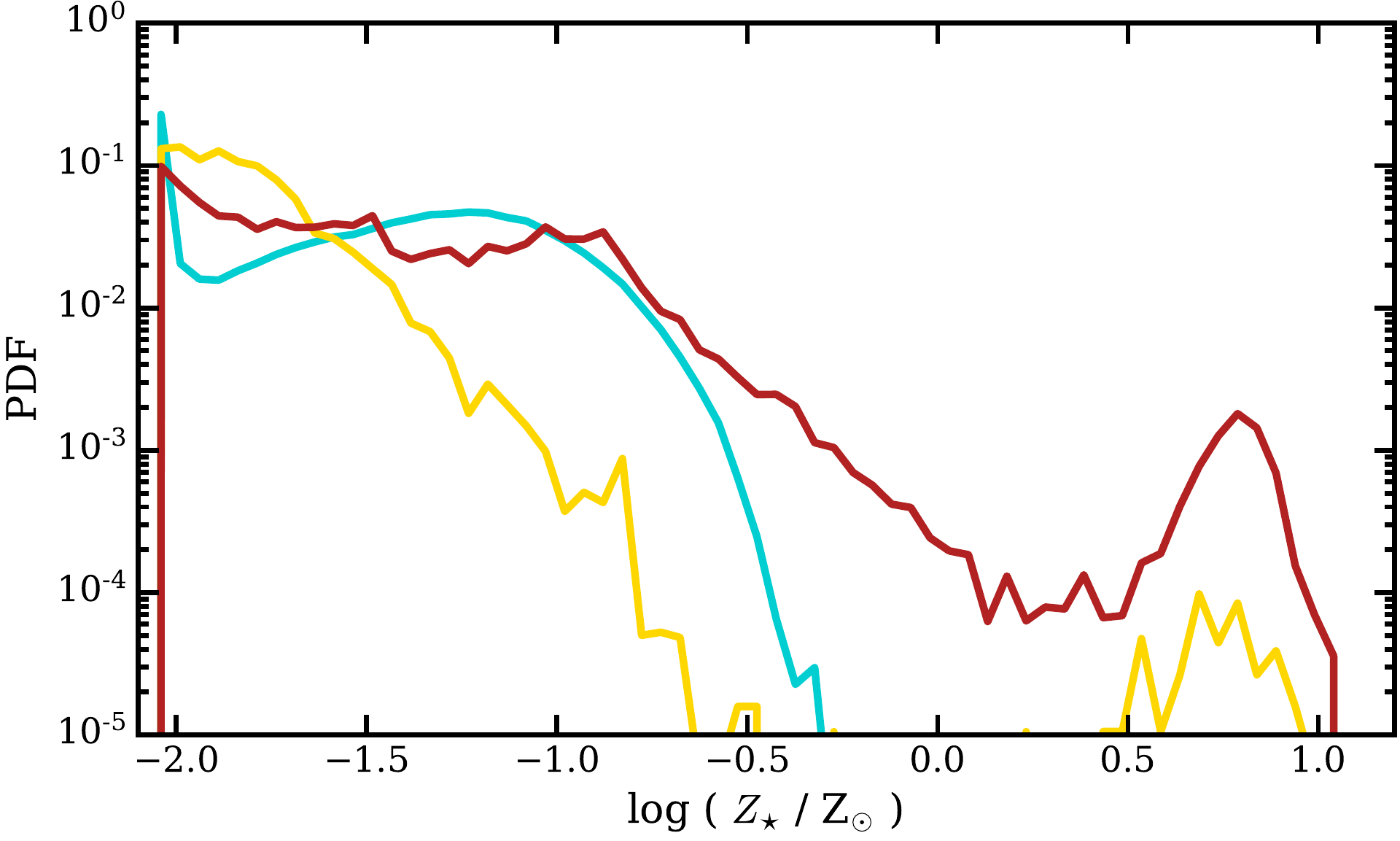}
  \caption[]{Metallicity distribution function of gas cells (\textit{upper panel}) and stars (\textit{lower panel}) in our various isolated runs. \fixSN has a lower average gas metallicity because more metals are expelled earlier. Due to this and stronger metal mixing, the stellar metallicity does not establish the same bimodal feature as exhibited by \varSN.}
  \label{fig:MDF_gas_ideal}
\end{figure}
\subsection{Metallicity}
The metals produced during the simulation are critical to the behaviour of the resulting galaxy. The gas cooling rates are strongly dependent on the metallicity, making enriched gas ($Z/\Zsun \sim 1$) cool at a rate an order of magnitude higher than less enriched gas ($Z/\Zsun \sim 0.1$) at the same density. Stars forming from dense cores inherit the metallicity of their cloud. A star's birth metallicity (with its birth mass) determines the star's lifetime. The higher the metallicity of a SN progenitor star, the shorter its life. In the case of AGB progenitor stars (in our model these are all stars with initial masses below 8\Msun, but without the stars in the unresolved part of the IMF) this trend is reversed. However the AGB metal returns are increased for higher metallicity stars. Thus, overall metal production is a run-away process where more metal-rich stars produce more metals in a shorter amount of time.

Fig.~\ref{fig:MDF_gas_ideal} shows the metallicity distribution function of all the gas cells within the simulation volume. Since we begin to evolve these idealized initial conditions at a metallicity of $1\%$ solar, there are no cells with a metallicity lower than this value. With no AGB returns and no SN returns, the distribution would remain a Dirac delta function at $1\%$ solar. So all higher metallicities in \nofb (blue line) are due to AGB returns. We note that this is a logarithmic plot, so that the small differences visible between \varSN (dark red line) and \fixSN (yellow line) runs are actually significant. {\varSN peaks at around $Z/\Zsun<-0.8$ with a tail to higher values. The entire distribution below a metallicity of around log $Z/\Zsun=-0.5$ is shifted to lower values for \fixSN. Due to the increased SN energy, many more metals escape the disk. }

The strong effect induced by differences in the model can be seen in the metallicity distribution of stars formed. This is shown in the lower panel of Fig.~\ref{fig:MDF_gas_ideal}. 
{Most stars in \varSN have log $Z/\Zsun<-1.0$, but the distribution exhibits a long, extended tail to higher metallicity with a second peak at super-solar values.} This tail is not (or almost not) present in \fixSN.
{It is produced by the high mass stars that do not inject energy but deposit large amount of metals in the gas. This highly enriched gas quickly cools and is turned into stars.} When there is equal energy injection for every massive star, the injected metals are more quickly blown out and do not collapse to form these highly enriched stars. The mean (log-averaged) stellar metallicity is $-1.45$ for \varSN. {With a final stellar mass of $\sim3.5\times10^7\,\Msun$, this easily matches the mass-metallicity relation within the scatter.} The mean stellar metallicities are $-1.80$ and $-1.47$ for the runs \fixSN and \nofb, respectively.
\begin{figure}
  \includegraphics[width=\linewidth]{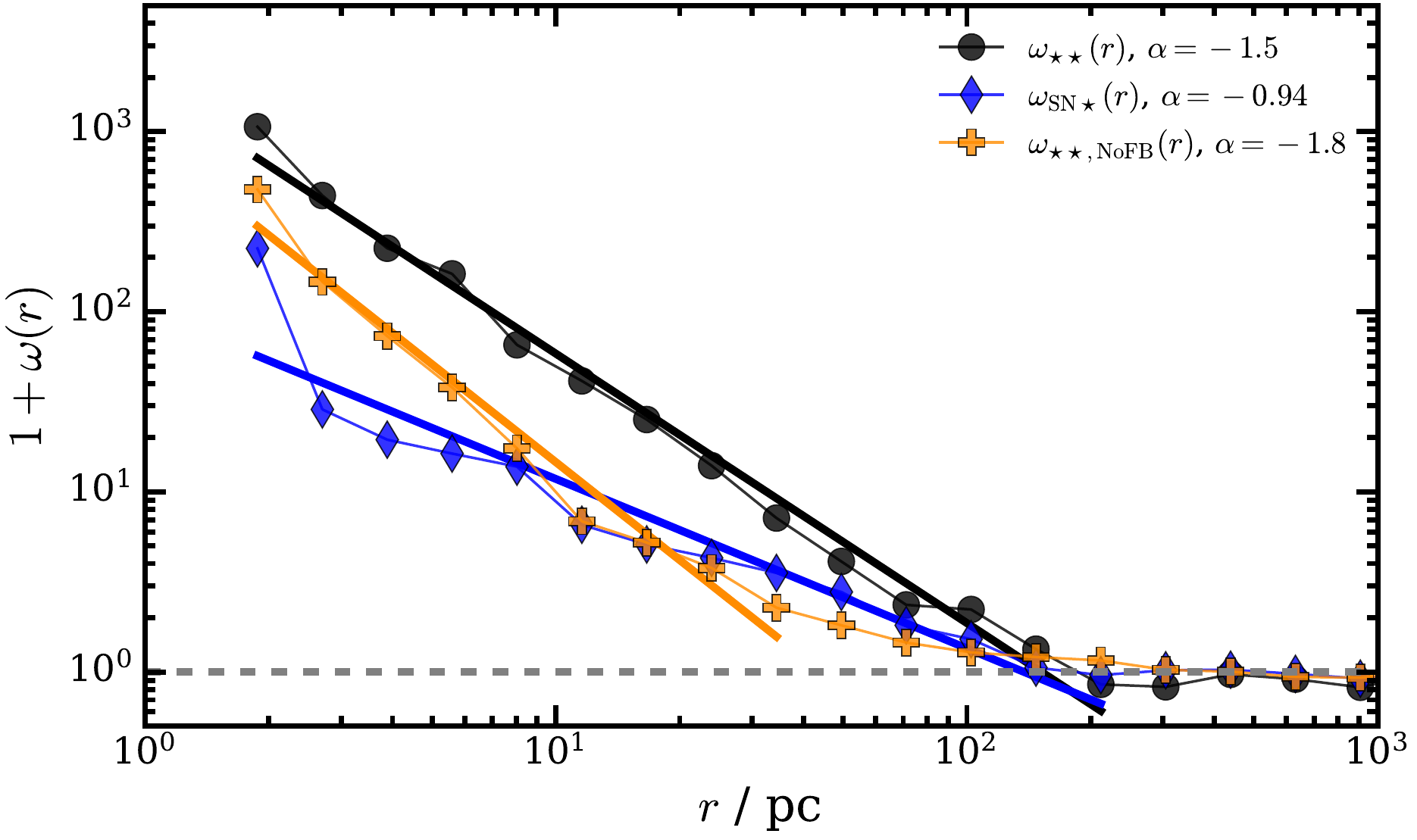}
  \caption[]{Two point correlation functions for young star particles in 2D, averaged over $40$\,Myr between $t=80-120$\,Myr. Black circles represent the auto-correlation function of young stars (age $< 1$\,Myr) in the \varSN model. Blue diamonds represent the cross-correlation of recent SN ($2\,{\rm Myr} > t_{\rm SN} > 1\,{\rm Myr}$) and young stars (age $< 1$\,Myr). The orange crosses represent the auto-correlation of young stars in \nofb. Thick lines with different colors show  power law fits for the different cases, whose slopes $\alpha$ are given in the legend. The black and blue fits were calculated between $1\,{\rm pc} < r < 200\,{\rm pc}$, while the orange line was fit for the range $1\,{\rm pc} < r < 30\,{\rm pc}$.}
  \label{fig:tpcf}
\end{figure}
\begin{figure*}
  \includegraphics[width=\linewidth]{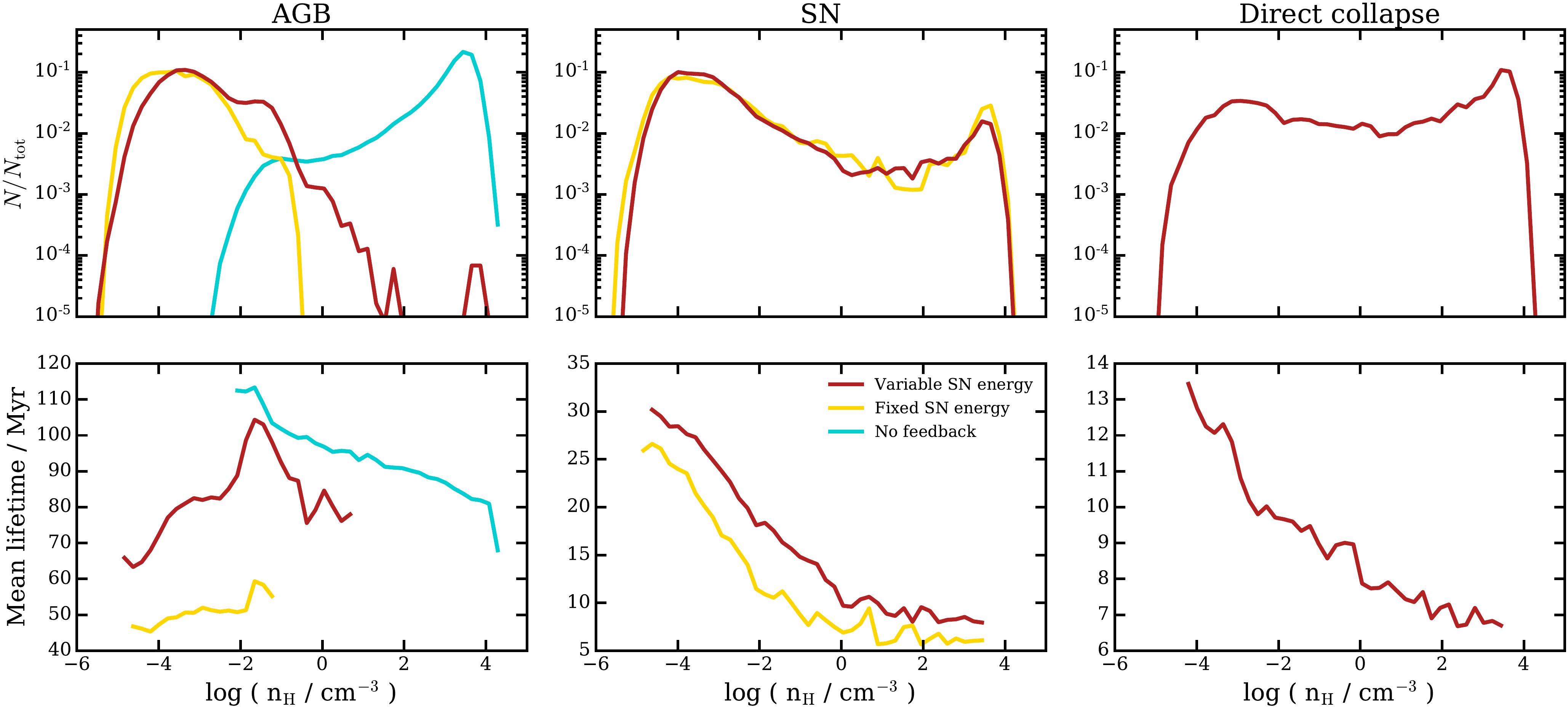}
  \caption[]{PDF of the gas density of cells (\textit{upper panels}) and the mean lifetime of stars binned by the cell density where the stars end their lives (\textit{lower panels}) and inject mass and metals (and energy in the case of the SN) in our various isolated runs, for AGB (\textit{left}), SN (\textit{center}) and direct collapse stars (\textit{right})  The shortest-lived SNe are exploding in gas at the SF threshold. However, the majority of the SNe occur in pre-evacuated bubbles at much lower densities of $\sim10^{-4}$\,\cc.}
  \label{fig:SN_density_ideal}
\end{figure*}
%
\subsection{Stellar and SN clustering}
We investigate the clustering of young stars by means of the projected two-point correlation function $\omega(r)$. We use the definition presented in \citet{Landy1993} to calculate the auto-correlation of young stars that formed in the last 1\,Myr, for which the distance $r$ is measured in two dimensions in the plane of the disk. To increase statistics, we sum the contribution over 100\,Myr, from $t=50-150$\,Myr where the SFR is highest. The black symbols in Fig.~\ref{fig:tpcf} show $\omega_{\star \star}(r)$ for young stars in \varSN. There is a clear clustering signal starting at $\sim150$\,pc. We fit this with a simple least squares power law and obtain a slope of $\alpha = -1.5$, as shown in the legend. \citet{Grasha2015} calculated the power-law slope of stellar clustering for the star-forming galaxy NGC 628 and obtain a mean slope of $-0.8$ with a break at $158$\,pc. Indeed, their youngest sub-sample (ages $< 10$\,Myr) exhibits a slope of $-1.51$. So, the clustering strength and extent of our simulated stars seems well in line with expectations from observations. 

We additionally compare the cross-correlation function $\omega_{{\rm SN} \star}(r)$ of recent SN ($2\,{\rm Myr} > t_{\rm SN} > 1\,{\rm Myr}$) with young stars ($< 1$\,Myr) to investigate how much a SN will affect the star formation in its vicinity in 1\,Myr. This is presented as the blue symbols and the corresponding fit in Fig.~\ref{fig:tpcf}. The break remains at approximately the same radius of $\sim150$\,pc. However, the slope is now strongly reduced to $-0.94$. This is a quantitative measure of how well SN feedback is able to suppress SF. While Fig.~\ref{fig:triggeredSF} hints at a possible positive feedback effect, where SN induce new SF at the edge of their bubbles, this cannot be directly seen in this cross-correlation measurement. Nevertheless, it also does not rule out a presence of this effect entirely. While the suppressive effect of SNe is clearly dominant, some induced SF could still be present, albeit as a weaker effect.

Lastly, in Fig.~\ref{fig:tpcf}, we also show the auto-correlation $\omega_{\star \star, {\rm NoFB}}(r)$ of young stars in \nofb in orange. Here we see much stronger clustering at smaller radii, with a slope of $-1.8$ and a break at $\sim30$\,pc. A comparison of the black and orange lines tells us that SNe tend to prevent the formation of small, tight clusters. Instead, the small-scale clustering of stars is reduced somewhat and spread out to larger distances. 
Since the stars form in clusters, this inevitably means the SN explosions happen in a clustered manner as well. This has profound repercussions for the effectiveness of feedback \citep[see e.g.][]{Gatto2015, Fielding2018, Hu2019}. As described in Section~\ref{sec:SNboxes}, the density into which a SN blastwave moves determines how well our model is able to resolve it. But more importantly, SN energy moving into a low density region will heat a larger volume and result in reduced SF, thus making the feedback effect more pronounced. 

The upper panels of Fig.~\ref{fig:SN_density_ideal} show the density distribution of the cells into which our dying stars inject mass, metals and, in the case of the SN, energy. The color coding is the same as in Fig.~\ref{fig:SFH_ideal}. The left panels show the distribution for the AGB returns, the central panel shows the SNe and the right panel shows the ``direct collapse'' stars, the massive stars that die without injecting energy. The density distribution is very similar in all runs with SNe. The AGB stars are injected into $\sim10^{-4}\,\cc$ gas. The SN also have a peak at $10^{-4}\,\cc$, but display a second smaller peak at $10^{3}-10^{4}\,\cc$. This broad distribution is consistent with the ``SN only'' run in \citetalias{Hu2017}, which is, however, somewhat less bi-modal. The direct collapse stars are reversed, their main peak is at high density but a small peak at low density is also present.

To help understand these distributions, we also show the mean lifetime of stars in each density bin in the lower panels. The direct collapse stars have the shortest lifetimes since they have the most massive progenitors. All runs with SN show a clear trend that the shortest lived massive stars die in the highest density, while the longer lived stars inject their mass in lower density regions. This trend is the same for the SN (lower, central panel). The AGB stars have a slightly different trend, since they live long enough that the effect of the SN bubbles originating from their birth environment is not felt any longer. 

So the most massive stars inject their feedback in gas that is very close to the density of star formation. Once the first SN explodes, an underdense bubble is produced that has an average density of $10^{-3}-10^{-4}\,\cc$. The longer lived stars increasingly inject their returns into this density rather than into gas that is close to the star formation threshold density.
The runs without SN have a strikingly different distribution. The AGB mass returns are primarily injected into gas that has a density close to our star formation threshold. This makes it apparent that the low density injection seen in the runs with SN is driven by SN clustering, or more accurately stellar clustering. For the runs without SNe the AGB lifetimes (blue and violet lines in the lower left panel) nevertheless exhibit a trend with density. This can be attributed to the fact that their lifetimes are very long relative to the SF cycle. This allows them to wander away from their birth environment, tending towards lower density given more time.
%
\subsection{Mass, metal and energy loading}
%
\begin{figure}
    \includegraphics[width=\linewidth]{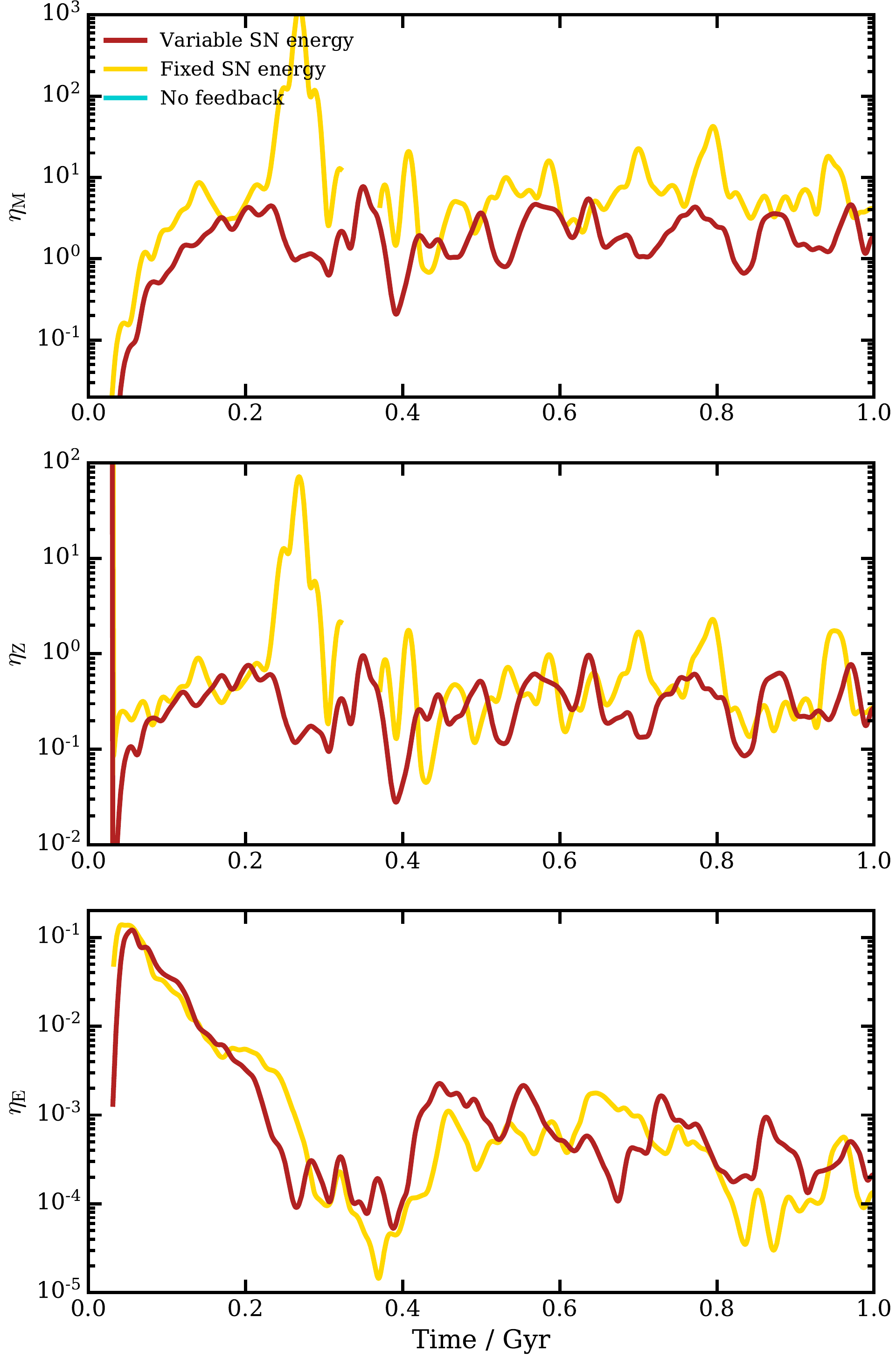}
  \caption[]{Mass, metal mass and energy loading over time in our various isolated runs. Runs without SNe have zero mass loading. The \varSN run shows a reduced mass loading with respect to \fixSN by a factor of $2-3$.}
  \label{fig:mass_loading_ideal}
\end{figure}
We now turn to an investigation of the outflows caused by our SNe.
The mass loading factor attempts to quantify the amount of outflowing mass relative to the amount of SF at a given time. While it is slightly arbitrary where and how the outflowing mass is measured, we choose to measure the mass of gas in cylindrical slabs above and below the disk placed at 2\,kpc distance from the plane of the disk. The radius of this cylindrical slab is set to the maximum radius of stars at that time. Its thickness is ${\rm d}L= 50\,$pc. We sum all mass $m_i$ of gas within this cylinder that has a positive $v_z$ velocity component (or negative in the case of the cylinder below the disk): 
\begin{equation}
    \dot{M}_{\mathrm{out}} = \frac{\sum m_i v_{z, i}}{{\rm d}L} .
\end{equation}
While this measures any local turbulence as additional mass loading, we stick to this definition for the sake of a fair comparison with other work such as \citeauthor{Hu2017}.
The metal loading, $\eta_{\rm M}$, is then computed as the outflow rate $\dot{M}_{\mathrm{out}}$ divided by the SFR. 
In the upper panel of Fig.~\ref{fig:mass_loading_ideal} we show this value as a function of time in the simulation. {\varSN exhibits mass loading factors averaging around $1-10$,} which is similar to the values reported in \citet{Hu2017,Hu2019}. \fixSN has a higher mass loading, reaching values of 1000. This is due to the overall more energetic feedback. Runs without energetic SN have zero mass loading.

In turn, the metal loading factor $\eta_{\rm Z}$ attempts to quantify the amount of metals ejected from the disk relative to the metals created by the SF process. 
As such, it can be computed as \citep[cf.][]{Emerick2019}:
\begin{equation}
    \eta_{\rm Z} = \frac{\dot{M}_{\mathrm{Z, out}}}{\mathrm{SFR}}\frac{M_{\star}}{M_{\mathrm{Z}}} ,
\end{equation}
where $M_{\star}$ is the stellar mass at each time and $M_{\mathrm{Z}}$ is the total metal mass created in the simulation at the corresponding time. We measure $\dot{M}_{\mathrm{Z, out}}$ in the same way as $\dot{M}_{\mathrm{out}}$, and show the evolution of this value in the central panel of Fig.~\ref{fig:mass_loading_ideal}. Metal loading factors are similar in all runs with SNe and increase throughout the simulation. The increase is in part due to the decreasing SFR. The values start at around 0.1 and increase to $\sim1$. This is not entirely unexpected.
\citetalias{Emerick2019} show metal loading factors for their $10^{9}$\,\Msun isolated dwarf, also finding values of order unity. Also, \citet{Hu2019} measure enrichment factors of $1-2$, {which is slightly higher than our values}.
\citet{Chisholm2018} use archival HST data to investigate the outflow properties of a small sample of galaxies and show that strongly metal enriched outflows shape the mass-metallicity relation for dwarf galaxies. They measure metal loading factors of $>100$ for two dwarf galaxies with stellar masses around $10^7\Msun$. However, they divide the outflow rate by the mean ISM metallicity instead of the produced metal mass, since this is not available from observations. Thus, this value is not directly comparable to the simulation definition.

The lower panel of Fig.~\ref{fig:mass_loading_ideal} shows the energy loading, $\eta_{\rm E}$. Consistent with the previous definitions of loading factors, this is defined as the energy outflow rate divided by the energy injection rate. We obtain a value of $\sim0.1$ at the beginning of the simulation, which then flattens to below $10^{-3}$ staying approximately constant until the end of 1\,Gyr.

\subsection{Phases of the ISM}
%
\begin{figure}
  \includegraphics[width=\linewidth]{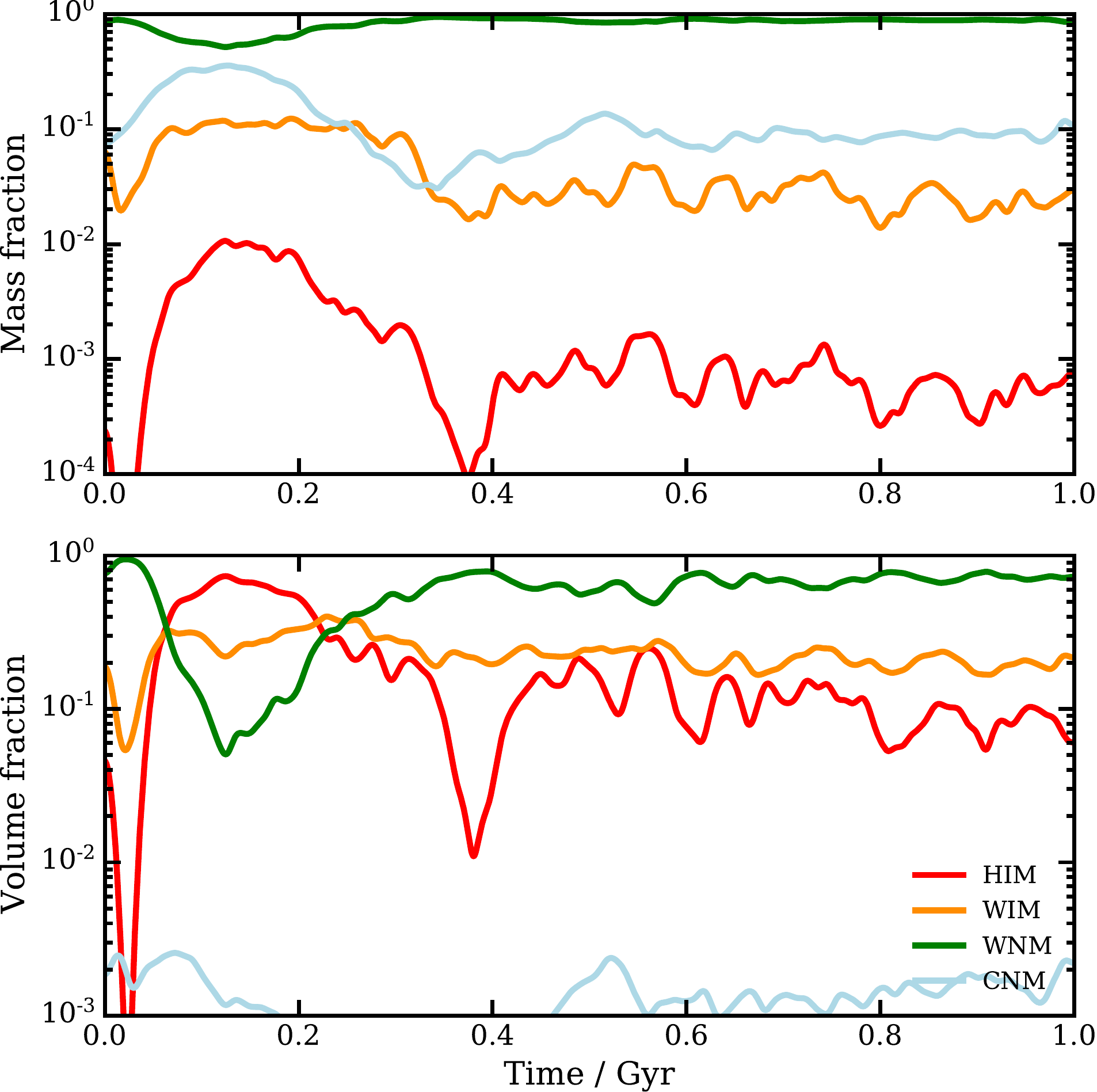}
  \caption[]{Fraction of gas within the disk in the phases HIM, WIM, WNM and CNM, as a function of time for the run \varSN. The upper panel shows the distribution weighted by the mass, while the lower panel applies a weighting by the volume instead. The disk is defined as a cylinder with height 200\,pc above the plane, and a radius given by the maximum radius of a star at that time. While the HIM makes up less than 1\% of the mass, {it occupies 10-20\% of the volume} of the ISM and is responsible for driving outflows.}
  \label{fig:him_wim_ideal}
\end{figure}
\begin{figure}
    \includegraphics[width=\linewidth]{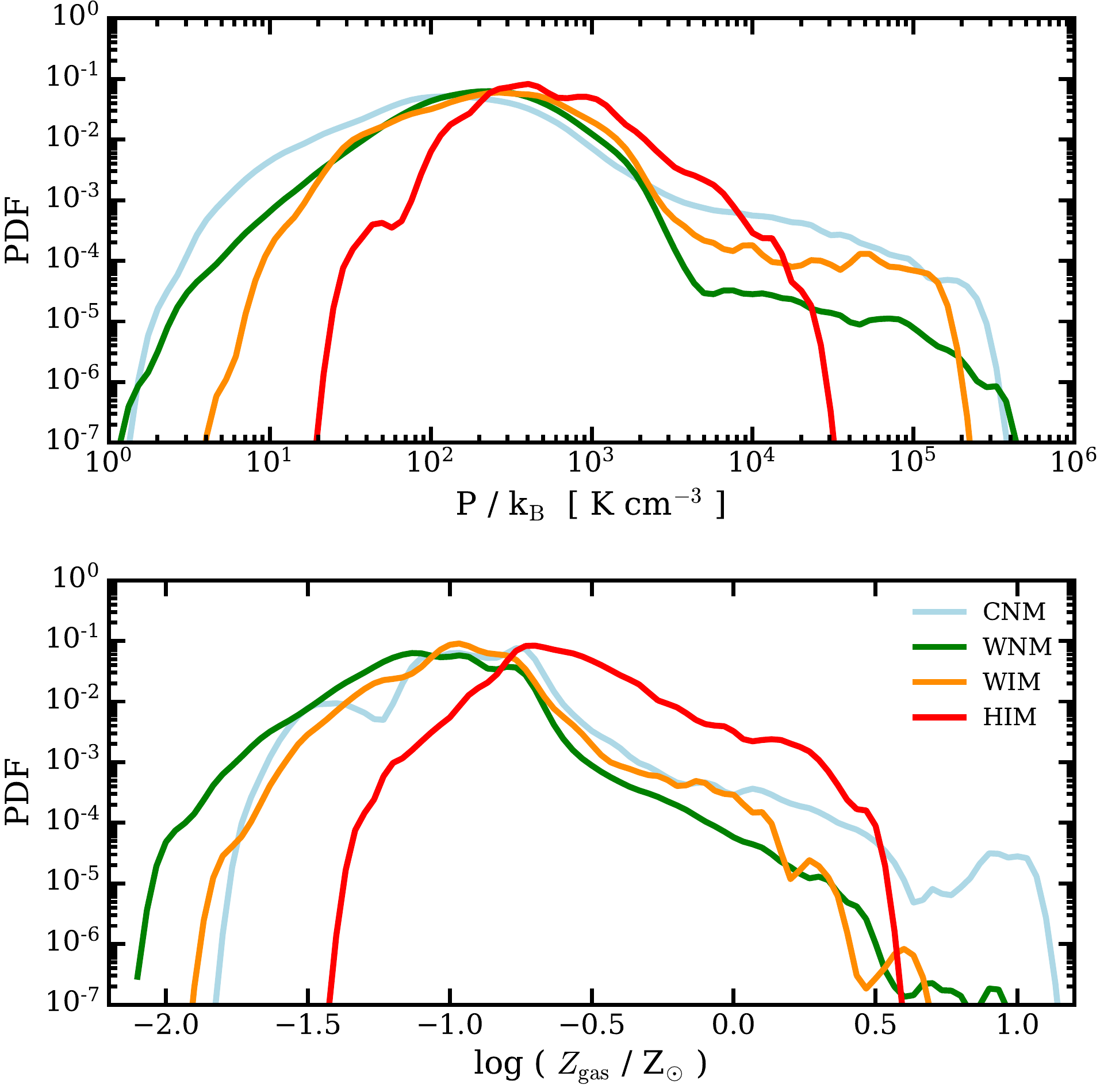}
  \caption[]{Pressure and metallicity PDF of each ISM phase within the disk after 1\,Gyr  for the \varSN model. The CNM, WNM and WIM are in pressure equilibrium. However, the CNM exhibits a tail to higher pressure that constitutes the majority of SF gas. The HIM is over-pressurized, reflecting the hot, expanding SN bubbles.}
  \label{fig:him_pressure_ideal}
\end{figure}
As \citet{MacLow1988} and others have shown, the primary driver of outflows is the hot gas ($\sim10^6$\,K) within the disk that is over-pressurized and blows bubbles of gas into the circum-galactic medium (CGM). The ISM is dominated in volume by hot gas in which individual cold, dense clumps sit. This is why momentum injection schemes for SN feedback are less well equipped to form a multiphase medium and reduce SF. In this section, we investigate the phase components of our ISM. As above, the ISM is defined within a cylinder centered on the disk with radius equal to the maximum stellar radius and height above the midplane of 200\,pc.

To better understand the structure of the ISM in our simulations, we split it into four phases: the hot ionized medium (HIM, $T>10^{5.5}$\,K), the warm ionized medium (WIM, $10^4\,{\mathrm{K}}<T<10^{5.5}\,{\mathrm{K}}$), the warm neutral medium (WNM, $10^2\,{\mathrm{K}}<T<10^{4}\,{\mathrm{K}}$), and finally the cold neutral medium (CNM, $T<10^2$\,K). Fig.~\ref{fig:him_wim_ideal} shows the mass and volume fractions of these four phases within the disk over time. The {WNM} comprises approximately 80-90\% of the mass at all times (after the burn-in phase) with most of the rest of the mass sitting in the {CNM} phase. {The volume is also dominated by the WNM with most of the rest of the volume in the WIM and around 10\% in the HIM}. If we turn to the top panel of Fig.~\ref{fig:him_pressure_ideal}, we can see the distribution of pressure within each phase. The bulk of CNM, WNM and WIM are in pressure equilibrium. However, there is a tail of CNM at higher pressures which constitutes most of the SF gas. The HIM has overall higher pressures. This is recently shock heated gas from SNe that is over-pressurized. {When this begins to cool, it forms the large tail to very high pressures of $>10^5$\,K\cc in the WIM and the WNM.}

Lastly, the lower panel of Fig.~\ref{fig:him_pressure_ideal} shows the metallicity distribution within each ISM phase.
The neutral media peak at just below a metallicity of -1. 
Whereas the HIM peaks at a value of -0.6, it has a long tail to very high metallicities well above solar. This phase is the first to be enriched by the SNe, since as we know from Fig.~\ref{fig:SN_density_ideal}, most SNe are injected in low density material. Since the SNe are injected with energy and metals simultaneously, it is not surprising that the metallicity in the hot phase is the highest. However, from Fig.~\ref{fig:him_wim_ideal} we also know that the CNM carries the most mass. Indeed, despite its lower overall metallicity, the majority of metal mass is locked up in the neutral medium. The ionized media hold a mere 2-3\% of the metal mass. The enrichment of the neutral media is slower and can be caused either by blastwaves penetrating cold clouds at their edges, or by hot enriched gas re-cooling into the colder phases.
\section{Discussion}
\label{sec:discussion}
%
\subsection{Comparison to other simulations}
\label{sec:comparisons}
While our new model is similar in nature to other simulation projects in the literature, we believe we have made some significant advancements. Here, we would like to highlight some of the differences to previous work.

As we have used the same initial conditions as \citet{Hu2016} and \citetalias{Hu2017}, we will compare to their results first. \citetalias{Hu2017} use the SPH code {\small GADGET-3} with the improved SPHGal formulation of SPH and set their mass resolution to $m_{\rm gas} = 4 \Msun$. A chemical network of \Htwo, H+ and CO allows for the formation of a multi-phase ISM. 
The authors also include radiation from young stars that is responsible for photo-heating. Cosmic rays and various chemical processes are followed to estimate the heating and cooling of the gas. This interstellar radiation field (ISRF) may initially reduce the SFR and later help maintain a steady rate.

The ``noPE-noPI-SN'' model of \citetalias{Hu2017} includes the most similar set of physics relative to our model. In this model, the SFH peaks at a value of around $10^{-3}\,\Msun \mathrm{yr}^{-1}$ and then flattens at around $2\times10^{-4}\,\Msun \mathrm{yr}^{-1}$. This is approximately an order of magnitude below our SFH. Our initial burst of SF is driven by triggered SF, which is not seen to this extent in \citetalias{Hu2017}.
There are a variety of model choices that differ and may account for the altered SFH. Possibly the largest difference is that \citetalias{Hu2017} use an SPH code while our model employs a moving mesh algorithm. This may cause the density peak of the shock within the SN shells to be more resolved with \arepo and therefore produce higher densities leading to more SF. 

We choose $0.5$\,pc softening for the gas and $10$\,pc for the dark matter. \citetalias{Hu2017} use $2$\,pc and $62$\, pc, respectively. When we test our model with the same softening values as \citetalias{Hu2017}, our SFR is reduced by more than a factor of two (see Fig. \ref{fig:SFH_tests}). Also, we set our SF threshold density to $10^3$\cc, while \citetalias{Hu2017} use a Jeans mass threshold where the mean density of SF is around an order of magnitude lower than ours. 
As such, the gas disk in our model is able to collapse to a thinner plane due to our smaller gas softening and it is given more time to collapse initially due to our higher SF density threshold. Thus, we get a higher initial burst in SF. The SN shock waves from this burst consequently trigger further SF in the adjacent clouds before the effects of negative feedback are able to set in. We do note that a further difference between the models is that the starting metallicity is $0.1\,Z_\odot$ in \citetalias{Hu2017} instead of our $0.01\,Z_\odot$. 

In general, however, our ISM structure is broadly consistent with the properties presented in \citetalias{Hu2017}. They report mass loading factors of 5-10, which is in the same range as ours. This is not surprising given that the hot gas fractions we find in the last 500\,Myr are also consistent with their values: their mass fraction in hot gas is around $2\times10^{-3}$, and the volume fraction is $0.05-0.1$. Similar hot gas fractions are  consequently expected to drive corresponding outflows. 
However, \citetalias{Hu2017}'s SNe injection density distribution peaks around $10^{-2}$\cc, which is two orders of magnitude higher than our finding. But since our SN produce triggered SF, we have a higher escape fraction and thus a lower gas surface density toward the end of the simulation, which may account for the difference. 
%
%

Another similar model is presented by \citetalias{Emerick2019}, who use the Enzo code to run an isolated dwarf galaxy (halo mass of $10^9\,\Msun$, one order of magnitude smaller than our isolated halo)
with feedback from individual stars. Their implementation
includes direct star-by-star modelling, stellar winds from massive and
AGB stars and an adaptive ray tracing method to include stellar
radiation. They also track non-equilibrium chemistry using
the Grackle code. Star formation proceeds stochastically and each cell produces
$\sim100 \Msun$ of stars stochastically drawn from the IMF. 
The formed stars are randomly placed within cell, which at the highest resolution level corresponds to a size of $1.8\,$pc.
Their feedback prescription drives large outflows with mass loading factors of $\sim50$ at $0.25\Rvir$, and $\sim10$ at \Rvir. They show that $\sim4\%$ of metals are retained in the disk, while $\sim50\%$ are ejected beyond \Rvir. Despite their smaller halo, these values are very similar to ours, with $\sim60\%$ of metals being blown beyond \Rvir in this model.
\citetalias{Emerick2019} also report hot phase (same as HIM in this work) mass fractions around $10^{-3}$ and volume fractions around 2-3\%.
Their density distribution of SN sites peaks at $10^{-3}$\,\cc in \citetalias[][see appendix C]{Emerick2019} which is 1 dex higher than what we find. We note that the SN sites are highly dependent on the star formation prescription and the SN injection mechanism. Thus, differences of this magnitude are not unexpected.

 Using the FIRE suite, \citet{Muratov2015} investigate the outflow properties in their simulations. For their smallest galaxies ``m09'' and ``m10'' they report mass loading factors of $100-1000$. This is significantly higher than the mass loading we measure. However, their simulations have a qualitatively different implementation of feedback, and are run at much lower resolution. An additional hindrance to a quantitative comparison is that our present simulation is based on an idealized, non-cosmological set up. {On the other hand, the stratified disk presented in \citet{Fielding2017} exhibits mass loading factors of around $0.1-1$, which is somewhat below our values. The authors show that mass loading is sensitive to resolution effects as well as to the extent of SN clustering.}
%
\citet{Smith2019} also report much higher SN injection densities for their simulation, peaking around $10^4$\,\cc. Since they use a thermal plus momentum injection scheme for their SN, it is plausible that they do not produce a volume filling hot phase within the ISM to drive outflows and reduce the gas density. This may explain their much higher SN injection density distribution.
Note that our SN density is based only on the cell itself, not a region of injection. This is different from, for example, \citet{Kim2017} who consider the mean density in the region of the SN bubble for this quantity. However, our density is nevertheless well below the majority of their SNe that expand into $10^{-2}-10^{-1}$\,\cc within the stratified disks they present. 
%

\subsection{Missing physical processes}
This paper discusses the first version of the LYRA model. As such, it includes only a subset of the physical processes known to be relevant for the scales we resolve. There are various models in the literature that have investigated the addition of further physics such as magnetic fields \citep[e.g.][]{Pakmor2013, Girichidis2018b}, cosmic rays \citep[e.g.][]{Simpson2016, Girichidis2018a, Chan2019}, interstellar radiation \citep[e.g.][]{Hopkins2011, Agertz2013, Roskar2014, Rosdahl2013, Peters2017} or a more detailed chemistry network \citep[e.g.][]{Glover2010}. We consider LYRA model as a new  sophisticated testing ground for many such modules and forthcoming papers will investigate these. Here, we  summarize our expectations on how the missing processes may affect our present conclusions. 

We expect magnetic fields to have an effect on the ISM clouds, stabilizing them against gravitational collapse, and thus reducing the SFR. Also cosmic rays that provide an additional heating source will influence when and how clouds cool. 

Binary stars are not included in any form at present but will affect stellar yields and stellar lifetimes. Information about binarity could be used for an effective white dwarf formation model that could in turn be used to model the SNIa rate. The explicit addition of SNIa will affect the iron abundance and increase the overall metallicity. We leave an investigation of these options for future work.

Radiation from young stars may arguably be the most critical ingredient missing from the present model. This can have a profound effect on the temperature distribution of the ISM \citep[e.g.][]{Peters2017, Emerick2018, Agertz2020} and, thus, its capability to cool efficiently. The SFR with radiation is likely to decrease by up to an order of magnitude. The ISM is expected to harbour more gas with a temperature around $10^4$\,K, since radiation provides a continuous heating source directly within the ISM. Additionally, the radiation from young stars is able to pre-evacuate dense regions of star formation before the first SNe explode, thereby reducing the density of SN sites and increasing the feedback efficiency of the explosions. 

{We do not explicitly include dust formation and destruction in the gas. However, the low temperature cooling rates were calculated assuming the ``Orion'' dust model in CLOUDY. According to \citet{Popping2017}, for dwarf galaxies with stellar masses below $10^7\,\Msun$ the dominant mode of dust formation is condensation in SN remnants. Including dust explicitly may decrease the gas phase metallicities and thus possibly also the resulting stellar metallicities, since dust formation depletes the gas.}

Photo-electric (PE) heating, i.e. the heating of dust by ultraviolet photons, is a feedback process that acts on the dense ISM directly, since the relevant photons mainly originate from young stars that are still embedded in their natal gas clouds. \citetalias{Hu2017} have shown that PE heating in conjunction with SNe can suppress star formation by a additional factor of 2-3, relative to the effect of SNe alone. The SNe are nevertheless required, since PE heating alone is unable to develop outflows and produce a multi-phase ISM. This is a much weaker suppression than reported in \citet{Forbes2016}, however the authors discuss the reasons for this in detail.
H17 have shown that  PE heating in conjunction with SNe can suppress star formation by an additional factor of 2-3, relative to the effect of SNe alone. 

A metallicity dependency of SN yields is not yet implemented in this model. SN yields are not well constrained, however some models such as \citet{Woosley1995} predict lower total metal yields for SN progenitor stars that have lower metallicity. This effect can be up to a factor of two in returned metal mass if some metals are present in the progenitor.
Fully primordial stars are not well understood, but SN models predict even less metal return from them. This effect may be especially important for cosmological simulations that begin with gas of primordial composition. Overall, we expect that an inclusion of  metallicity dependent yields would decrease the total metal mass produced. This, in turn, could increase cooling times and reduce SF.

Furthermore, we do not include a stellar wind model that accounts for mechanical momentum injection by young stars. Such a model is expected to further decrease the ambient density of SN sites since the wind acts before the first SNe go off. This may further increase the effectiveness of SNe driving outflows. A stellar wind model coupled to LYRA will be presented in a forthcoming paper.

\section{Conclusion}
\label{sec:conclusion}
We have presented the new LYRA model for galaxy formation, which is the first of its kind to include variable SN energy. Individual stars sampled from the IMF are traced as individual collisionless particles and follow an initial mass-dependent stellar evolution pathway. The radiative cooling of gas is modelled down to 10\,K and includes a prescription for molecular hydrogen. SN events are traced individually and expel thermal energy, mass and metals according to their progenitor mass. AGB stars, as well as the more massive direct collapse stars, return mass and metals to the ISM.  

We have demonstrated that at the densities where the majority of our SNe explode, we resolve the Sedov phase and reproduce the expected amount of kinetic energy based on a purely thermal injection. This makes sub-grid momentum boost prescriptions to account for resolution limitations superfluous, thereby paving the way for fully realistic treatments of the multi-phase ISM in cosmological simulations of galaxy formation.
We validate this new model on an isolated dwarf galaxy simulated with a very high target gas mass resolution of 4\,\Msun. The gravitational softening for both gas and stars is 0.5\,pc in these runs. \\

Our main results are as follows:
\begin{itemize}
\item 
We find that our SN feedback prescription self-consistently produces a hot phase within the ISM that drives massive outflows, reducing gas density and suppressing SF.
\item
 Metals are efficiently ejected from the halo through these winds, reducing the overall metallicity to observed values.
\item The metallicity distribution is significantly affected by the energy injection scheme. The model that accounts for direct collapse stars (without energy injection) produces a tail of high metallicity stars. This tail is otherwise not created because the expulsion of the metals by SNe is more efficient.
\item
A model with variable SN energy yields an altered outflow behaviour with mass loading factors of $1-10$, which is lower by a factor of 2-3 with respect to corresponding fixed energy model. 
\item
The variable energy model expels 8\% of its gas beyond the virial radius, as opposed to nearly 12\% in the fixed energy model.
\item
The variable energy model expels $\sim40\%$ of the metals it produces beyond the virial radius, while the fixed models expel 60-70\%. {The higher loss of metals is related to the chimneys being wider with fixed energy.} 
\item
 Clustered SF plays a major role in the effectiveness of feedback. Since SF clustering leads directly to SN clustering, it ultimately significantly enhances the efficiency of the SN in heating gas and driving winds, because the majority of explosions then occur in low density material.
 \item
 Our results are well in line with previous studies in the literature, especially with regard to the ISM phase structure and outflow behaviour of the gas. Our SFR is high in comparison to previous results, which may point toward the need for the inclusion of further feedback processes such as radiation from young stars.
 \item {Stellar evolution couples directly to the feedback. This is ignored in integrated feedback prescriptions, making them insensitive to the spatially distributed energy deposition. }
\end{itemize}
The obvious next goal with LYRA is to run cosmological zoom-in simulations of field dwarf galaxies at similarly high mass resolution as presented here. These calculations will be presented in a forthcoming paper. Additionally, we are developing LYRA further, adding more physics such as magnetic fields and radiation from stars. While resolving more physics introduces further complications, it is necessary to take this step to arrive at truly faithful models of the ISM, and to begin to better understand how the small scale details of SN feedback can affect integrated galaxy properties across cosmic times. Such models also offer the prospect to inform the development of more reliable sub-grid models for galaxy formation that can be used to carry out studies of the formation of full galaxy populations at much lower resolution and in very large volumes.

%
\section*{Acknowledgments} 
The authors thank the anonymous referee who read our draft with much attention and significantly contributed to the improvement of this publication. We also thank Enrico Garaldi and Simona Vegetti for useful discussions. TN acknowledges
support from the Deutsche Forschungsgemeinschaft (DFG, German
Research Foundation) under Germany’s Excellence Strategy –
EXC-2094 – 390783311 from the DFG Cluster of Excellence
‘ORIGINS’.
This research was carried out on the High Performance Computing resources of the {\small FREYA} cluster at the Max Planck Computing and Data Facility (MPCDF) in Garching operated by the Max Planck Society (MPG).
\section*{Data availability}
The data underlying this article were accessed from the High Performance Computing resources of the Max Planck Computing and Data Facility (MPCDF, \url{https://www.mpcdf.mpg.de}). The derived data generated in this research will be shared on reasonable request to the corresponding author.
%
\renewcommand{\refname}{REFERENCES}
\bibliographystyle{mnras}
\bibliography{bib}
%
\appendix
\counterwithin{figure}{section}
\section{Robustness and resolution tests}
\label{sec:appendix}

\begin{figure}
  \includegraphics[width=0.95\linewidth]{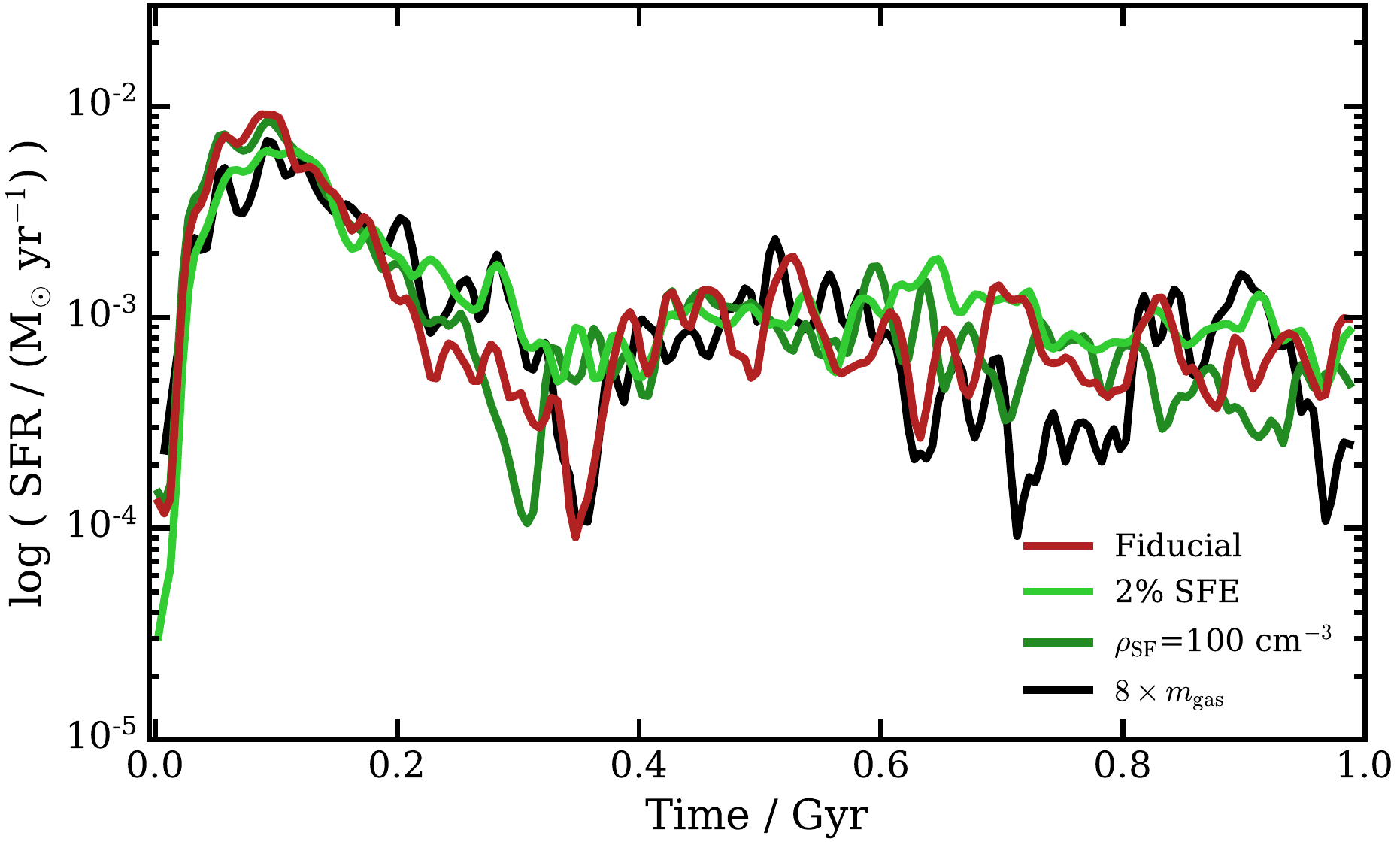}
  \caption[]{Star formation history in various test runs that modify different numerical model parameters.}
  \label{fig:SFH_tests}
\end{figure}

\begin{figure}
  \includegraphics[width=0.95\linewidth]{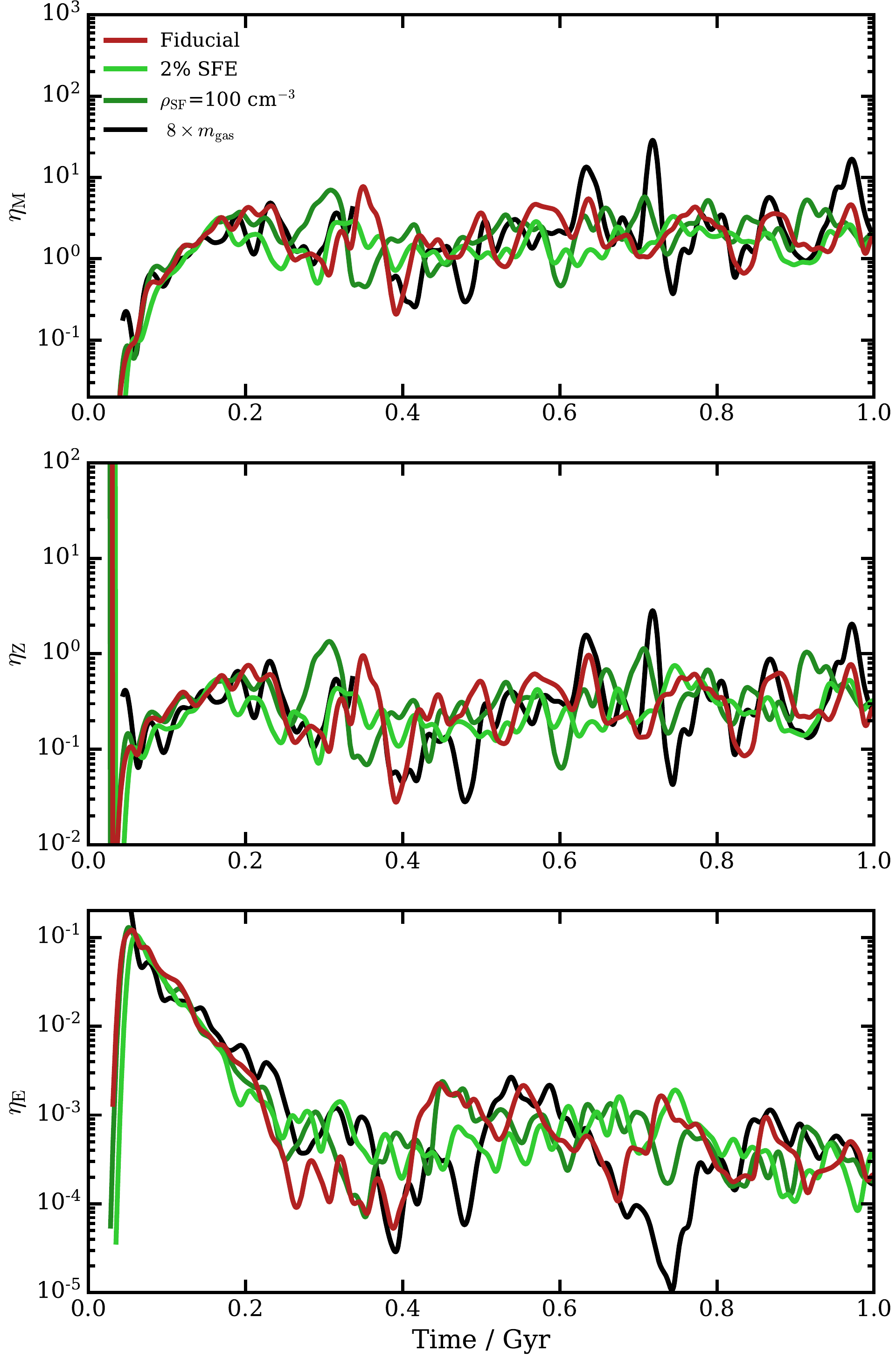}
  \caption[]{Mass, metal mass and energy loading of a set of test runs that vary different numerical model parameters.
  \label{fig:loading_tests}}
\end{figure}


We here discuss { three} test simulations, varying certain parameters or aspects of our model:
\begin{enumerate}
    \item 
 $\rho_{\rm SF} = 100\,\cc$ means we have set the density threshold for SF to 100\,\cc instead of $10^3$\,\cc. 
\item
 2\% SFE: We don't change the SF efficiency above $10^4$\,\cc.
\item
 $8 \times m_{\rm gas}$: We increase the target gas mass by a factor of 8 from 4\,\Msun to 32\,\Msun.
\end{enumerate}

Fig.~\ref{fig:SFH_tests} shows the SFH of the fiducial run and all three test runs. Neither the SF efficiency, the density threshold of star formation nor a reduced mass resolution significantly alter the SFH beyond its intrinsic scatter. 
%

\begin{figure}
    \includegraphics[width=0.95\linewidth]{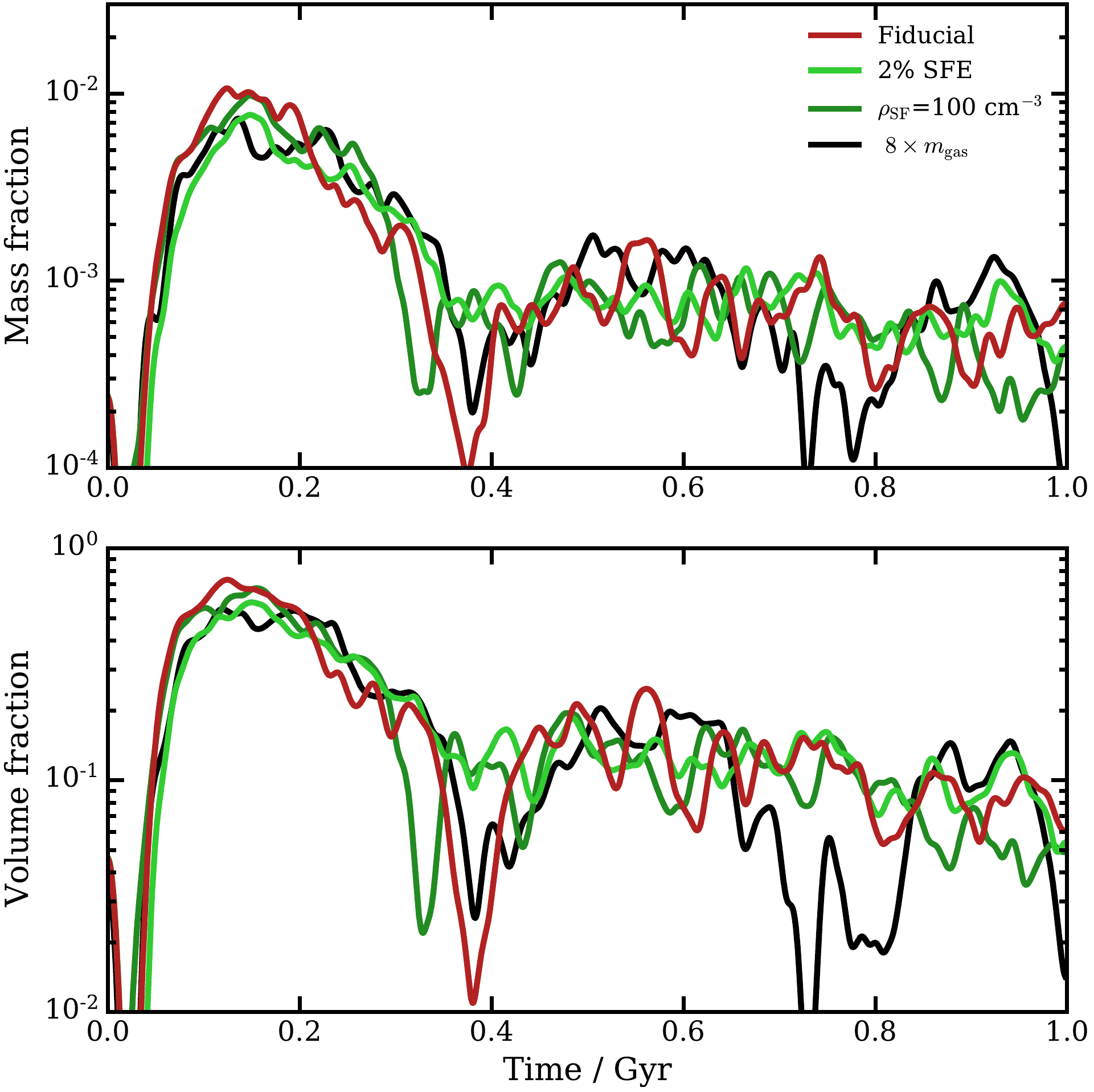}
  \caption[]{Mass and volume fractions of the HIM phase of a set of test runs that vary different numerical model parameters.}
  \label{fig:him_tests}
\end{figure}

Fig.~\ref{fig:loading_tests} shows that the mass and metal loading are insensitive to our parameter choices. Additionally, we show the mass and volume fraction of the HIM phase in Fig. \ref{fig:him_tests}. 
\\
\label{lastpage}
\bsp
%
%
\end{document}